\documentclass[]{aa}

\usepackage{color }
\usepackage{natbib}
\usepackage{amsmath}
\usepackage{amssymb}
\usepackage{graphicx}

\newcommand{\bl}[1]{\mbox{\boldmath$ #1 $}}

\title{Thermal evolution of protoplanetary disks: from $\beta$-cooling to decoupled gas and dust temperatures}
\titlerunning{Thermal evolution of protoplanetary disks}
\author{Eduard I. Vorobyov\inst{1,2}, Ryoki Matsukoba\inst{3}, Kazuyuki Omukai\inst{3}, and Manuel Guedel\inst{1}} 
\authorrunning{Vorobyov et al.}
   
\institute{ 
    University of Vienna, Department of Astrophysics, Vienna, 1180, Austria \\
    \email{eduard.vorobiev@univie.ac.at} 
    \and 
    Ural Federal University, 51 Lenin Str., 620051 Ekaterinburg, Russia
    \and
    Astronomical Institute, Graduate School of Sciences, Tohoku University, Aoba, Sendai, Miyagi 980-8578, Japan
    }


\date{}

\begin{document}

 \abstract
   {}
   {We explore the long-term evolution of young protoplanetary disks with different approaches to computing the thermal structure determined by various cooling and heating processes in the disk and its surroundings. }
   {Numerical hydrodynamics simulations in the thin-disk limit were complemented with three thermal evolution schemes: a simplified $\beta$-cooling approach  with and without irradiation, in which the rate of disk cooling is proportional to the local dynamical time, a fiducial model with equal dust and gas temperatures calculated taking viscous heating, irradiation, and radiative cooling into account, and also a more sophisticated approach allowing  decoupled dust and gas temperatures.  }
   {We found that the gas temperature may significantly exceed that of dust in the outer regions of young disks thanks  to additional compressional heating caused by the infalling envelope material in the early stages of disk evolution and slow collisional exchange of energy between gas and dust in low-density disk regions.
   The outer envelope however shows an inverse trend with the gas temperatures dropping below that of dust. The global disk evolution is only weakly sensitive to temperature decoupling. Nevertheless, separate dust and gas temperatures may affect the chemical composition, dust evolution, and disk mass estimates. Constant-$\beta$ models without stellar and background irradiation fail to reproduce the disk evolution with more sophisticated thermal schemes because of intrinsically variable nature of the $\beta$-parameter. Constant-$\beta$ models with irradiation can better match the dynamical and thermal evolution, but the agreement is still incomplete. }
   {Models allowing separate dust and gas temperatures are needed when emphasis is placed on the chemical or dust evolution in protoplanetary disks, particularly in sub-solar metallicity environments. }

\maketitle

\section{Introduction}

Protoplanetary disks are an important ingredient of star and planet formation. They form during the gravitational contraction of rotating pre-stellar cloud cores thanks to the conservation of angular momentum of the infalling material. It is considered that most of the core material is processed by the disk before it lands on the growing protostar or leaves the system through protostellar jets and outflows. This processing includes alterations in the chemical composition and growth of sub-micron particles to cm-sized pebbles, providing building blocks for planets. These processes critically depend on the thermal balance in the disk and knowing the properties of protoplanetary disks is therefore of prime importance for our understanding of star and planet formation.

The studies of protoplanetary disks have traditionally followed two separate pathways. Global simulations follow the collapse of pre-stellar cores to the protostellar stage characterized by the formation of a star and protostellar disk \citep[e.g.,][]{Machida2010,Joos2013,Seifried2013,Tsukamoto2015}. The forming disks usually show a very complex behaviour depending on the mass of the core, amount of initial rotation in the core, and the strength of magnetic fields \citep[e.g.][]{Bate2018,Wurster2019}. The interaction with the environment in the form of jets, outflows, and infalling material makes the interpretation of numerical simulations a challenging task.  An alternative approach is to look into the evolution of isolated, so to say, already-formed disks \citep[e.g.,][]{Kley1999,Boss2002,Stamatellos2007,Mayer2007}. Such an approach allows focusing on particular aspects of disk evolution and usually permits a better numerical resolution, making this approach particularly valuable for our understanding of the subtleties of star and planet formation.

One of such important aspects of disk evolution is gravitational instability and fragmentation, which is considered to be a possible gateway for the formation of giant planets and brown dwarfs \citep[e.g.][]{Boss2002,Mayer2007,Boley2010,Vorobyov2013,Meru2015,Nayakshin2017, Mercer2017}.
Gravitational instability and fragmentation are particularly sensitive to the disk mass, but also to the thermal balance in the disk controlled largely by disk cooling, viscous and stellar heating.
Starting from \citet{Gammie2001}, it has become increasingly popular to employ the so-called $\beta$-parameterization to describe the cooling processes in the disk when studying the disk propensity to gravitational fragmentation \citep[e.g.,][]{Rice2003,Cossins2009,Meru2011,Boss2017,Deng2017}. In this approach, the rate of disk cooling is parameterized in terms of the $\beta$-parameter, which is the product of the local cooling time $t_{\rm c}$ and local angular velocity $\Omega$. The popularity of this approach can be explained by its simplicity, which allows avoiding  a complicated physics and numerics often involved with solving for the full energy balance equation and at the same time permits obtaining valuable insights into an important physical process -- disk gravitational fragmentation. 

The often used approach is to adopt a uniform $\beta$-parameter throughout the disk, but vary its value to determine its effect on disk gravitational fragmentation. There is however no justification that the $\beta$-parameter should be uniform throughout the disk. Moreover, it is not clear if the $\beta$-parameterization describes accurately the thermal balance in the disk. Several numerical studies addressed the disk propensity to fragment depending on whether the simplified $\beta$-cooling or a more sophisticated cooling-heating scheme were used \citep[e.g.,][]{Gammie2001,Gammie2003}, but a systematic study of the temperature distribution in the disk on global evolutionary timescales has not been performed yet.  
It is therefore important to understand the risks that are involved with using the simplified $\beta$-approximation. 

In this paper, we consider three different approaches to describing the thermal balance in the disk. First, we consider the scheme that takes radiative cooling, stellar and viscous heating, and PdV work into account in the limit of equal dust and gas temperatures. This approach has been extensively used in one- and-two dimensional disk dynamics simulations \citep{Gammie2003,Rice2009,VB2010,Zhu2012} and also adapted to three-dimensional smoothed-particle simulations \citep{Stamatellos2007}. Second, we introduce a new cooling-heating scheme that allows a separate calculation of the gas and dust temperatures in protoplanetary disks. This scheme is similar in methodology (but not exactly the same) to the methods  earlier presented by \citet{Pavlyuchenkov2015} and \citet{Bate2015}. We particularly search for disk regions where the gas temperature can deviate notably from that of dust. Finally, we consider the simplified $\beta$-cooling and determine the applicability of this simplified approach in describing disk evolution.  

We use disk models in the thin-disk limit to study the effect of different cooling-heating schemes. A simplified disk dynamics allows us to focus on the thermal properties of the disk and run disk simulations for a much longer time than in full three-dimensional simulations. Nevertheless, we believe that our results regarding the applicability of the $\beta$-approximation and importance of separate dust and gas thermal evolution  will remain valid in fully three-dimensional disk models. More importantly, our thermal model can find applications in simulations of low-metallicity disks,  where decoupling of dust and gas temperatures is expected to be significant.

The paper is organized as follows. In Sect. \ref{model} we describe in detail different cooling-heating schemes that we used in our disk models. 
In Sect.~\ref{newmodel} we describe the disk evolution using the cooling-heating scheme with separate dust and gas temperatures.
In Sect.~\ref{oldmodel} we compare the disk evolution in models with separate and similar dust and gas temperatures. In Sect.~\ref{BetaCooling} we consider models with a simplified $\beta$-cooling. Main results are summarized in Sect.~\ref{Conclusions}.

\section{Model description}
\label{model}
In this section, we describe the main aspects of our model regarding the gas dynamics computations, while the subsequent subsections elaborate on computations of the thermal balance in the disk. 
We use numerical hydrodynamics simulations in the thin-disk limit to compute the formation and
global evolution of young circumstellar disks. To avoid too small time steps, we set a dynamically inactive sink cell in the center of our computational domain with a radius of $r_{\mathrm{sc}}=5$~au.

The starting point of each simulation is the gravitational collapse of a pre-stellar core.
In the adopted thin-disk approximation, the core has the form of a flattened pseudo-disk, a spatial configuration that can be expected in the presence of rotation and large-scale magnetic fields \citep[e.g.,][]{Basu1997}. As the collapse proceeds, the inner regions of the core spin up and a  centrifugally balanced circumstellar disk forms when the inner infalling layers of the core hit the centrifugal barrier near the sink cell. The material that has passed to the sink before the instance of circumstellar disk formation constitutes a seed for the central star, which  grows further through accretion from the circumstellar disk.
The infalling core continues to land at the outer edge of the circumstellar disk until the core depletes. The infall rates on the circumstellar disk are in agreement with what can be expected from the free-fall collapse \citep{Vorobyov2010}. Computations continue up to 0.5~Myr, thus covering the entire embedded phase and the early T Tauri phase of disk evolution.

We take into account turbulent viscosity described via 
the Shakura \& Sunyaev $\alpha$-parameterization and disk self-gravity. 
The forming protostar is not just a source of gravity. Its characteristics, such as the radius and photospheric luminosity, are calculated in line with the disk evolution using the stellar evolution tracks obtained with the STELLAR code \citep{Yorke2008}. These characteristics are further used to calculate the total stellar luminosity and the radiation flux impinging the surface of the disk and contributing to its heating in models where detailed disk cooling and heating are taken into account.

The equations of mass and momentum in the thin-disk limit are:
\begin{equation}
\frac{\partial\Sigma}{\partial t}=-\bl{\nabla}_{p}\cdot\left(\Sigma \bl{v}_{p}\right),\label{eq:mass}
\end{equation}

\begin{equation}
\frac{\partial}{\partial t}\left(\Sigma \bl{v}_{p}\right)+\left[\bl{\nabla} \cdot \left(\Sigma 
\bl{v}_{p}\otimes \bl{v}_{p}\right)\right]_{p}=-\bl{\nabla}_{p} P+\Sigma \bl{g}_{p}+\left(\bl{\nabla}\cdot
\bl{\Pi}\right)_{p},\label{eq:momentum}
\end{equation}
where subscripts $p$ and $p'$ refer to the planar components $(r,\phi)$
in polar coordinates, $\Sigma$ is the gas mass surface density, $P$ is the vertically integrated
gas pressure calculated via the ideal equation of state as $P=(\gamma-1)e$, $\gamma$ is the ratio of specific heats, $\bl{v}_{p}=v_{\mathrm{r}}\hat{\bl{r}}+v_{\mathrm{\phi}}\hat{\bl{\phi}}$
is the velocity in the disk plane, $\bl{g}_{p}=g_{\mathrm{r}} \hat{\bl{r}} 
+ g_{\mathrm{\phi}} \hat{\bl{\phi}}$
is the gravitational acceleration in the disk plane (including that of the disk and the star) and 
$\bl{\nabla}_{\mathrm{p}}=\hat{\bl{r}}\partial/\partial r+\hat{\bl{\phi}}r^{-1}\partial/\partial\phi$
is the gradient along the planar coordinates of the disk. 
Turbulent viscosity enters the basic equations via the viscous stress
tensor $\bl{\Pi}$ and we calculate the magnitude of kinematic viscosity $\nu$ using the 
$\alpha$-parameterization with a spatially uniform $\alpha$-parameter. Two limiting cases were considered: an MRI-active disk with $\alpha=0.01$ and an MRI-suppressed disk with $\alpha=10^{-4}$.

\subsection{$\beta$-cooling}
In its simplest form, the energy balance equation can be described as follows
\begin{equation}
\frac{\partial e}{\partial t}+\bl{\nabla}_{\mathrm{p}}\cdot\left(e v_{p}\right)=-P\left(\bl{\nabla}_{p} \cdot v_{p}\right)- {e\over t_{\rm c}} + \left(\bl{\nabla} \cdot v\right)_{pp'}: \bl{\Pi}_{pp'},
\label{eq:energy1}
\end{equation}
where  $e$ is the internal energy of gas per surface area.  The characteristic cooling time is related to the $\beta$-parameter as $t_{\rm c}=\beta / \Omega$. This approach, referred to further in the text as the $\beta$-cooling scheme,  takes into account the advection of internal energy with the gas flow, heating/cooling through adiabatic compression/expansion of gas flows (first term on the right-hand side), radiative cooling approximated by the $\beta$-parameter, and viscous heating (last term on the right-hand side). We note that in many studies with the $\beta$-cooling scheme, but not in our paper,  viscous heating is neglected. To avoid a catastrophic overcooling of the disk we turn off the $\beta$-cooling term as soon as the gas temperature drops below a threshold value set equal to 4~K. This beta-cooling approach was employed in studies of disk fragmentation in, for example,  \citet{Rice2003,Meru2011,Boss2017,Rice2018}. In our study,
the energy equation~(\ref{eq:energy1}) is closed with the ideal equation of state $P=e (\gamma-1)$ for a perfect gas, for which the ratio of specific heats is set equal to a constant value of $\gamma=1.4$.

 One drawback of the above approach is that it does not take stellar irradiation into account, although this heating mechanism can be important once the star has formed. There are various modifications to the standard $\beta$-cooling scheme \citep[see, e.g.,][]{Baehr2015}, but in this study we adopt the following form 
\begin{equation}
\frac{\partial e}{\partial t}+\bl{\nabla}_{\mathrm{p}}\cdot\left(e v_{p}\right)=-P\left(\bl{\nabla}_{p} \cdot v_{p}\right)- {e-e_{\rm irr}\over t_{\rm c}} + \left(\bl{\nabla} \cdot v\right)_{pp'}: \bl{\Pi}_{pp'},
\label{eq:energyIrr}
\end{equation}
where $e_{\rm irr}$ is the internal energy per surface area defined exclusively by stellar and background irradiation as
$e_{\rm irr} = \Sigma {\cal R} T_{\rm irr} / \mu$, with the mean molecular weight set equal to $\mu=2.33$.
Here, $T_{\rm irr}$ is the irradiation temperature calculated as 
\begin{equation}
T_{\rm irr}^4=T_{\rm bg}^4+\frac{F_{\rm irr}(r)}{\sigma},
\label{fluxCS}
\end{equation}
where $T_{\rm bg}$ is the uniform background temperature set equal to the initial temperature of the natal cloud core and $F_{\rm irr}(r)$ is the radiation flux  absorbed by the disk surface at a radial distance  $r$ from the
central star. The flux is calculated as 
\begin{equation}
F_{\rm irr}(r)= \frac{L_\ast}{4\pi r^2} \cos{\gamma_{\rm irr}},
\label{fluxF}
\end{equation}
where $\gamma_{\rm irr}$ is the incidence angle of radiation arriving at the disk surface (with respect to the normal) at radial distance $r$. The incidence angle is calculated using a flaring disk surface as described
in \citet{VB2010}. The stellar luminosity $L_\ast$ is the sum of the accretion and stellar photospheric luminosities.
The modified $\beta$-cooling term works as a relaxation process with a timescale $t_{\rm c}$ towards the thermal state defined by $T_{\rm irr}$. The stronger the mismatch between $e$ and $e_{\rm irr}$, the faster the system strives to attain the thermal state defined by irradiation (for a fixed value of $\beta$). We note that the other terms on the right-hand side of Equation~(\ref{eq:energyIrr}) can act to push the thermal balance away from that established by stellar and background irradiation.

\subsection{Similar thermal evolution of gas and dust}
\label{THES1}
A more sophisticated approach to computing the thermal balance in the disk involves solving the energy equation considering the effects of viscous heating, disk radiative cooling and stellar heating via irradiation. The pertaining equation reads as
\begin{equation}
\frac{\partial e}{\partial t}+\bl{\nabla}_{\mathrm{p}}\cdot\left(e v_{p}\right)=-P\left(\bl{\nabla}_{p} \cdot v_{p}\right)-\Lambda+\Gamma+\left(\bl{\nabla} \cdot v\right)_{pp'}: \bl{\Pi}_{pp'},
\label{eq:energy2}
\end{equation}
where $\Lambda$ and $\Gamma$ are the cooling and heating rates due to dust cooling and stellar (and background) irradiation, respectively.

This approach, referred to further in the text as the thermal evolution scheme~1 (or ThES1), was employed to study disk fragmentation in, for example, \citet{Gammie2003,VB2010,Zhu2012}. The difference between these studies lied in the degree of sophistication in calculating the radiative cooling and disk heating, in neglecting or taking viscous heating into account. For instance, \citet{Gammie2003} considered only a cooling term and neglected disk heating through stellar irradiation and viscosity. Their cooling term read as 
\begin{equation}
    \Lambda = {16 \over 3} \sigma T_{\rm mp}^4 {\tau_{\rm R} \over 1+ \tau_{\rm R}^2},
    \label{coolterm}
\end{equation}
where $\tau_{\rm R}$ is the mean Rosseland optical depth, $T_{\rm mp}$ is the midplane temperature of dust (and gas), and $\sigma$ is the Stefan-Boltzmann constant. \citet{Zhu2012} added irradiation heating by the central star (and also background) in the form
\begin{equation}
    \Gamma = {16 \over 3} \sigma T_{\rm irr}^4  {\tau_{\rm R} \over 1+ \tau_{\rm R}^2}.
    \label{heatterm}
\end{equation}
\citet{VB2010} also considered  viscous heating due to turbulence via $\alpha$-parameterization.

The form of the cooling-heating terms may vary depending on the degree of sophistication in calculating radiative cooling of dust from the disk surface. In this work we use the expression derived in \citet{Dong2016}
\begin{equation}
\Lambda=\frac{8\tau_{\rm P} \sigma T_{\rm mp}^4}  {1+2\tau_{\rm P} + 
{3 \over 2}\tau_{\rm R}\tau_{\rm P}},
\end{equation}
\begin{equation}
\Gamma=\frac{8\tau_{\rm P} \sigma T_{\rm irr}^4 }{1+2\tau_{\rm P} + {3 \over 2}\tau_{\rm R}\tau_{\rm
P}},
\end{equation}
where $\tau_{\rm P}$ is the Planck optical depth. We note that the cooling and heating rates in \citet{Dong2016}
were written for one side of the disk and need to be multiplied by a factor of 2. The energy equation~(\ref{eq:energy2}) is closed with the ideal equation of state $P=e (\gamma-1)$, where ratio of specific heats is set equal to a constant value of $\gamma=1.4$. We also note that the cooling and heating terms of the form similar to Equations~(\ref{coolterm}) and (\ref{heatterm}) are often used together with the viscous equation 
of \citet{Pringle1981} for the gas surface density to compute the thermal balance in the disk \citep[e.g.,][]{Rice2010,Kimura2016}.

\subsection{Different thermal evolution of gas and dust}
\label{ThES2}
The thermal evolution scheme considered in Sect.~\ref{THES1} makes no difference between the gas and dust temperatures. This is a valid approximation at high densities when collisions between gas molecules and dust particles are sufficiently frequent to establish a thermal equilibrium between these two disk subsystems on timescales much shorter than the dynamical one. However, it is not clear a priori if this condition is fulfilled throughout the entire extent of a protostellar/protoplanetary disk. 
In the outer disk regions densities may be too low to provide strict thermal coupling between gas and dust. 
In this section we present a new cooling-heating scheme, referred to as the thermal evolution scheme 2 or ThES2, which is designed to lift the limitation of equal gas and dust temperatures. In ThES2 we also have a spatially and temporally varying ratio of specific heats $\gamma$, thus lifting another limitation of the $\beta$-cooling and ThES1 schemes, for which a perfect gas with constant $\gamma$ was assumed.

In ThES2 we do not make a clear distinction between the cooling and heating rates as was done with $\Lambda$ and $\Gamma$ in the previous sections and introduce the integrated rate of energy loss or gain per surface area $Q_{\rm tot}$. The evolution equation for the gas internal energy per surface area in ThES2 reads as
\begin{equation}
\frac{\partial e}{\partial t}+\bl{\nabla}_{\mathrm{p}}\cdot\left(e v_{p}\right)=-P\left(\bl{\nabla}_{p} \cdot v_{p}\right) - Q_{\rm tot} + \left(\bl{\nabla} \cdot v\right)_{pp'}: \bl{\Pi}_{pp'},
\label{eq:energy3}
\end{equation}
where $Q_{\rm tot}$ is defined as
\begin{equation}
    Q_{\rm tot} = \left(Q_{\rm cont}+Q_{\rm H2}+Q_{\rm HD} + Q_{\rm chem} +Q_{\rm metal}\right) 2 H,
    \label{CoolingAll}
\end{equation}
where $H$ is the vertical scale height calculated assuming a local hydrostatic equilibrium in the gravitational field of the star and disk \citep[see][]{VB2009}, $Q_{\mathrm{cont}}$ is the rate of radiative energy loss (or gain) in the infrared continuum, $Q_{\mathrm{H_{2}}}$ is the H$_{2}$ line cooling rate, $Q_{\mathrm{HD}}$ is the HD line cooling rate, $Q_{\mathrm{chem}}$ is the chemical cooling/heating rate (through chemical reactions)), and $Q_{\mathrm{metal}}$ is the metal line cooling/heating rate.  All constituents of $Q_{\rm tot}$ are volumetric cooling or heating rates and, for simplicity, they are assumed to be independent of the vertical distance from the disk midplane. This assumption allows us to convert the volumentric rates to the rates per surface area by means of vertical integration and multiplication by the disk thickness 2$H$. We describe how to calculate these individual cooling rates below. 

The net rate of continuum cooling by energy transport from gas to radiation per unit volume is
\begin{equation}
    Q_{\mathrm{cont}} = 4\pi\left( \eta-\chi_{\mathrm{a}}J \right)~,
    \label{Eq:cnt_rate}
\end{equation}
where $\eta$ is the emission coefficient, $J$ is the mean intensity
and $\chi_{\mathrm{a}}$ is the absorption coefficient, given by 
\begin{equation}
    \chi_{\mathrm{a}} 
    = (\kappa_{\mathrm{P,d}} + \kappa_{\mathrm{P,g}})\rho~,
\end{equation}
with the mass density $\rho$. We calculate the Planck mean opacities using the tables from \cite{Semenov2003} for the dust and
\cite{Mayer2005} for the gas.  We note that Semenov opacities are defined per unit gas mass assuming a dust-to-gas mass ratio of 1:100.
The emission coefficient is 
\begin{equation}
    \eta = \frac{\sigma\rho}{\pi}\left( \kappa_{\mathrm{P,g}}T_{\mathrm{g}}^{4} + 
       \kappa_{\mathrm{P,d}}T_{\mathrm{d}}^{4} \right)~,
\end{equation}
where $T_{\mathrm{g}}$ and $T_{\mathrm{d}}$ are the gas and dust temperatures, respectively. The gas temperature is determined from the ideal equation of state ${P}=\Sigma {\cal R} T_{\rm g}/ \mu$, where $\mu$ is the mean molecular weight, $\sigma$ is the Stefan-Boltzmann constant,  and $\cal{R}$ is the universal gas constant.

The dust temperature is determined  in the steady-state limit by the energy balance on dust grains due to the thermal emission, 
absorption, and collision with gas \citep{Omukai2010}:
\begin{equation}
    \kappa_{\mathrm{P,d}}B(T_{\mathrm{d}}) = \kappa_{\mathrm{P,d}}J + \Gamma_{\mathrm{coll}}~,
    \label{Eq:dust_balance}
\end{equation}
where $B(T)$ is the Planck function, given by 
\begin{equation}
    B(T) = \frac{\sigma}{\pi}T^{4}~,
\end{equation}
where $T$ is the temperature of dust or stellar irradiation (see Eq.~\ref{irrad} below)  and $\Gamma_{\mathrm{coll}}$ is the heating rate of dust through collisions with gas particles. 
The collisional heating rate is \citep{Hollenbach1979}
\begin{equation}
    \Gamma_{\mathrm{coll}} = 4.4\times 10^{-6}\left( f/\rho \right)_{\mathrm{dust}}n_{\mathrm{H}}
        \left( \frac{T_{\mathrm{g}}}{1000~\mathrm{K}} \right)^{1/2} \left( T_{\mathrm{g}}-T_{\mathrm{d}} \right)~,
        \label{Eq:ColHeat}
\end{equation}
where $\left(f/\rho \right)_{\mathrm{dust}}$ is the total volume of dust per unit gas mass and $f$ is the mass fraction of dust grains, both are taken from \cite{Pollack1994}.  We note that the steady-state assumption for the dust temperature allowed us to eliminate $\Gamma_{\rm coll}$ from  the gas internal energy equation by rewriting $\Gamma_{\rm coll}$ in terms of the Planck function and the mean intensity.

The mean intensity used in Equations (\ref{Eq:cnt_rate}) and (\ref{Eq:dust_balance}) is 
\begin{equation}
    J = \frac{1}{1+x}\left( B(T_{\mathrm{irr}}) + x\frac{\eta}{\chi_{\mathrm{a}}} \right)~,
    \label{irrad}
\end{equation}
where $x$ is the function that smoothly connects the optically thin and thick limits, written by \cite{Tanaka2014} as 
\begin{equation}
    x = \tau_{\mathrm{P}} + \frac{3}{4}\tau_{\mathrm{P}}\tau_{\mathrm{R}}~,
\end{equation}
with the Planck and Rosselanck mean optical depths $\tau_{\mathrm{P}}$ and $\tau_{\mathrm{R}}$, respectively.
The Planck (Rosseland) mean optical depth is calculated as 
\begin{equation}
    \tau_{\mathrm{P(or~R)}} = \frac{1}{2}\left( \kappa_{\mathrm{P(or~R),d}}+\kappa_{\mathrm{P(or~R),g}} \right)\Sigma~,
\end{equation}
We obtain the Rosseland mean opacities in the similar manner to the Planck mean opacities.

The H$_{2}$ and HD-line cooling rates are calculated by  the following similar form:
\begin{equation}
    Q_{\mathrm{H_{2}(HD)}} = \overline{\beta}_{\mathrm{esc,H_{2}(HD)}}
       Q_{\mathrm{H_{2}(HD),thin}}\mathrm{e}^{-\sqrt{\tau_{\mathrm{P}}\tau_{\mathrm{R}}}}~,
\end{equation}
where $Q_{\mathrm{H_{2}(HD),thin}}$ is the cooling rate in the optically thin regime given by the fitting function 
for H$_{2}$ from \cite{Glover2015} and for HD from \cite{Flower2000}.
We take into account the line-averaged escape probability $\overline{\beta}$ to consider the effect of photon trapping 
in large column density case.
The values of line-averaged escape probabilities for H$_{2}$ and HD are obtained 
by using the fitting functions in \cite{Fukushima2018} and Equation~(\ref{Eq:HD_escape}) in Appendix \ref{App:HD-escape}.

The chemical cooling/heating are the processes associated with chemical reactions. We follow the chemical evolution of 8 species, H, H$_{2}$, H$^{+}$, H$^{-}$, D, HD, D$^{+}$, and e, and take into account 21 hydrogen and 6 deuterium reactions summarized in Table~\ref{Table:reactions}.
We consider H ionization/recombination and H$_{2}$ dissociation/formation as the chemical cooling/heating processes. 
The chemical cooling rate is 
\begin{equation}
    Q_{\mathrm{chem}} = \left( \epsilon_{\mathrm{H}}\frac{\mathrm{d}y(\mathrm{H}^{+})}{\mathrm{d}t} 
       - \epsilon_{\mathrm{H}_{2}}\frac{\mathrm{d}y(\mathrm{H}_{2})}{\mathrm{d}t}\right)n_{\mathrm{H}}~,
\end{equation}
where $\epsilon_{\mathrm{H}}$=13.6 eV and $\epsilon_{\mathrm{H}_{2}}$=4.48 eV are the binding energies. 
The chemical fraction of species $i$ is defined 
using the number density of species $i$, $n(i)$, and that of hydrogen nuclei $n_{\mathrm{H}}$ as follows:
\begin{equation}
    y(i) = \frac{n(i)}{n_{\mathrm{H}}}~.
\end{equation}
The number density of hydrogen nuclei is 
\begin{equation}
    n_{\mathrm{H}} = \frac{\rho}{(1+4y_{\mathrm{He}})m_{\mathrm{H}}}
\end{equation}
where $y_{\mathrm{He}}$ is the number fraction of He relative to hydrogen nuclei and 
$m_{\mathrm{H}}$ is the hydrogen nuclei mass. 

We consider the atomic fine-structure line emission of CII and OI as the metal line cooling $Q_{\rm metal}$. 
We model CII as a two level system and OI as a three level system and
count level populations from the statistical balance among each level. We take the level energies, 
the spontaneous radiative decay rates, and the collisional deexcitation rate coefficients from \cite{Hollenbach1989}. 
The metal line cooling/heating rate can be divided into the line cooling/heating rates of CII and OI as follows:
\begin{equation}
    Q_{\rm metal} = Q_{\rm CII} + Q_{\rm OI}~,
\end{equation}
\begin{eqnarray}
    Q_{\rm CII(OI)} &=& y_{\rm CII(OI)}n_{\rm H}\left(Z/Z_{\rm local}\right) \notag \\
    \displaystyle &\times&\sum_{\mathrm{ul}} h\nu_{\rm ul}\beta_{\rm esc,ul}A_{\rm ul}f_{\rm u}\frac{S(\nu_{\rm ul})-B(\nu_{\rm ul};T_{\rm rad})}{S(\nu_{\rm ul})}~,
\end{eqnarray}
where the chemical fractions of CII and OI are $y_{\rm CII} = 9.27\times10^{-5}$ and $y_{\rm OI} = 3.568\times10^{-4}$, 
$Z/Z_{\rm local}$ is the metallicity relative to solar one, 
$h\nu_{\rm ul}$ is the energy difference between the upper level u and the lower level l, 
$\beta_{\rm esc,ul}$ is the line escape probability, 
$A_{\rm ul}$ is the spontaneous radiative decay rate, 
$f_{\rm u}$ is the occupancy of upper level, 
$S(\nu_{\rm ul})$ is the source function, and 
$B(\nu_{\rm ul};T_{\rm rad})$ is the Planck function. 
We note that the metal lines heat gas if the gas temperature is lower than the irradiation temperature.
In this work, $Z/Z_{\rm local}$ is unity. 
The line escape probability is 
\begin{equation}
    \beta_{\rm esc,ul} = \left( \frac{1-\mathrm{e}^{-\tau_{\rm ul}}}{\tau_{\rm ul}} \right)\mathrm{e}^{-\sqrt{\tau_{\rm P}\tau_{\rm R}}}~, 
\end{equation}
where the optical depth for line emission $\tau_{\rm ul}$ is given by 
\begin{equation}
    \tau_{\rm ul} = \frac{c^{3}}{8\pi^{3/2}\nu_{\rm ul}^{3}}A_{\rm ul}\left( \frac{g_{\rm u}}{g_{\rm l}}f_{\rm u}-f_{\rm l} \right)\frac{N_{\rm{CII(OI)}}}{v_{\rm th}}~,
    \label{Eq:line-tau}
\end{equation}
where $g_{\rm u,l}$ is the statistical weight of upper level u and lower level l, 
$N_{\rm CII(OI)}$ is the column density of CII (OI), and 
$v_{\rm th}$ is the thermal velocity.
The column density of CII (OI) is 
\begin{equation}
    N_{\rm CII(OI)} = 2Hn_{\rm H}y_{\rm CII(OI)}~.
\end{equation}
The thermal velocity is 
\begin{equation}
    v_{\rm th} = \sqrt{\frac{2k_{\rm B}T_{\rm g}}{\mu m_{\rm H}}}~,
    \label{Eq:thermal-velocity}
\end{equation}
where $k_{\rm B}$ is the Boltzmann constant. 
The source function is calculated by 
\begin{equation}
    S(\nu_{\rm ul}) = \frac{2h\nu_{\rm ul}^{3}}{c^{2}}\left[\frac{g_{\rm u}f_{\rm l}}{g_{\rm l}f_{ \rm u}}-1\right]^{-1}~.
\end{equation}

 Our thermal model is based on the minimum model of \citet{Omukai2005}, which include only CII and OI line cooling (without solving C and O chemistry) and dust cooling
in addition to the primordial gas thermal and chemical processes. This model can reproduce the temperature evolution calculated by more elaborate models relatively well.
Note also that line cooling is only important at low densities ($\leq 10^4$~cm$^{-3}$).

\subsection{Initial and boundary conditions }

\begin{table*}
\center
\caption{\label{table1}Model parameters}
\begin{tabular}{cccccccc}
 &  &  &  &  &  &  &       \tabularnewline
\hline 
\hline 
Model & $M_{\mathrm{core}}$ & $\beta_{\rm c}$ &  $\Omega_{0}$ & $r_{0}$ & $\Sigma_{\rm g,0}$ & $r_{\mathrm{out}}$ & $\alpha$ \tabularnewline
 & [$M_{\odot}$] & [\%] &  [$\mathrm{km\,s^{-1}\,pc^{-1}}$] & [au] & [$\mathrm{g\,cm^{-2}}$] & [pc] \tabularnewline
\hline 
1v & 1.39 & 0.43 &  1.56 & 1560 & 0.09 & 0.075 & $10^{-2}$
\tabularnewline
2v & 0.64 & 0.46 &  1.73 & 1560 & 0.09 & 0.038 & $10^{-2}$
\tabularnewline
2 & 0.64 & 0.46 &  1.73 & 1560 & 0.09 & 0.038 & $10^{-4}$
\tabularnewline
3v & 0.28 & 0.98 &  2.8 & 1560 & 0.09 & 0.02 & $10^{-2}$
\tabularnewline
3 & 0.28 & 0.98 &  2.8 & 1560 & 0.09 & 0.02 & $10^{-4}$
\tabularnewline
\hline 
\end{tabular}
\center{ \textbf{Notes.} $M_{\mathrm{core}}$ is the initial core
mass, $\beta_{\rm c}$ is the ratio of rotational to gravitational energy of the core,  $\Omega_{0}$ and $\Sigma_{\rm g,0}$ are the angular velocity and gas surface density at the center of the core, $r_{0}$ is the radius
of the central plateau in the initial core, and $r_{\mathrm{out}}$ is the initial radius of the
core, and $\alpha$ is the value of the viscous $\alpha$-parameter.}
\end{table*}

In this work we considered five model cores, the parameters of which are provided in Table~\ref{table1}. 
The initial radial profile of the gas surface density $\Sigma$ and
angular velocity $\Omega$ of the pre-stellar core has the
following form: 
\begin{equation}
\Sigma=\frac{r_{0}\Sigma_{0}}{\sqrt{r^{2}+r_{0}^{2}}},\label{eq:sigma}
\end{equation}
\begin{equation}
\Omega=2\Omega_{0}\left(\frac{r_{0}}{r}\right)^{2}\left[\sqrt{1+\left(\frac{r}{r_{0}}\right)^{2}}-1\right],\label{eq:omega}
\end{equation}
where $\Sigma_{0}$ and $\Omega_{0}$ are the angular velocity and
gas surface density at the center of the core and $r_{0}$
is the radius of the central plateau. This radial profile
is typical of pre-stellar cores with a supercritical mass-to-flux ratio that are
formed through ambipolar diffusion, with the specific angular momentum
remaining constant during axially-symmetric core collapse \citep{Basu1997}. 
All pre-stellar cores are initially unstable to gravitational collapse, but differ in the amount of mass and angular momentum. In particular, model~1 is the most massive, while model~3 is the least massive one. Besides, model~3 is distinguished by a factor of 2 higher initial ratio of rotational to gravitational energy. The initial gas and dust temperatures are set equal to 10~K. 

The initial chemical composition of the cores in the ThES2 is as follows. 
We calculate the time evolution of the central density, temperature, and chemical composition of the collapsing cloud core 
with the one-zone treatment as in \cite{Omukai2005} until the central density reaches $10^{6}$ cm$^{-3}$.
The values of the chemical fractions of 8 species at that time are $y(\mathrm{H})=3\times10^{-10}$, $y(\mathrm{H}_{2})=0.5$, 
$y(\mathrm{H}^{+})=y(\mathrm{e})=10^{-8}$, $y(\mathrm{D})=2\times10^{-16}$, $y(\mathrm{HD})=3\times10^{-5}$, and $y(\mathrm{H}^{-})=y(\mathrm{D}^{+})=0$.

We distinguish between different cooling-heating schemes by adding the corresponding prefix. For example (ThES1)-model~1 would correspond to model~1 with the thermal evolution scheme 1. In addition, we put a letter ``v'' after the model number to denote the models with an increased value of the viscous $\alpha$-parameter, thus simulating a fully MRI-active disk.

The inner boundary condition located at $r_{\rm sc}$ should be chosen with a certain care. If the inner boundary allows for matter to flow only in
one direction from the active disk to the sink cell, then
any wave-like motions near the inner boundary, such as
those triggered by spiral density waves in the disk, would
result in a disproportionate flow through the sink-disk
interface. As a result, an artificial depression in the gas
density near the inner boundary develops in the course
of time because of the lack of compensating back flow
from the sink to the disk. A solution to this problem
was proposed in \citet{Vorobyov2018}, where a free inflow-outflow boundary condition was introduced, allowing matter to flow freely from the disk to the central sink cell and vice versa according to the computed mass transport rate through the sink-disk interface.  In particular, the mass of material $\Delta M_{\rm flow}$ (always positive definite) that passes through the sink-disk interface is further split into two components $\Delta M_\ast$  and  $\Delta M_{\rm s.c.}$, which are used to update the gas surface density in the sink cell $\Sigma_{\rm s.c.}$ and the stellar mass $M_\ast$ according to the following algorithm:
\begin{eqnarray}
 \mathrm{if}\,\, \Sigma_{\rm s.c.}^n < \overline{\Sigma}_{\rm in.disk}^n\,\, \mathrm{and}  \,\, v_r(r_{\rm s.c.})<0  \,\, \mathrm{then} \nonumber\\
 \Sigma_{\rm s.c.}^{n+1}=\Sigma_{\rm s.c.}^n &+& \Delta M_{\rm s.c.}/S_{\rm s.c.} \nonumber\\
 M_\ast^{n+1}&=&M_\ast^n+\Delta M_\ast \nonumber \\
 \mathrm{if}\,\, \Sigma_{\rm s.c.}^n < \overline{\Sigma}_{\rm in.disk}^n\,\, \mathrm{and}  \,\, v_r(r_{\rm s.c.})\ge 0  \,\, \mathrm{then} \nonumber\\
 \Sigma_{\rm s.c.}^{n+1}=\Sigma_{\rm s.c.}^n &-& \Delta M_{\rm flow}/S_{\rm s.c.} \nonumber\\
 M_\ast^{n+1}&=&M_\ast^n \nonumber \\
 \mathrm{if}\,\, \Sigma_{\rm s.c.}^n \ge \overline{\Sigma}_{\rm in.disk}^n\,\, \mathrm{and} \,\, v_r(r_{\rm s.c.})<0 \,\, \mathrm{then} \nonumber\\
 \Sigma_{\rm s.c.}^{n+1}&=& \Sigma_{\rm s.c.}^n \nonumber\\
 M_\ast^{n+1}&=& M_\ast^n + \Delta M_{\rm flow} \nonumber \\
  \mathrm{if}\,\, \Sigma_{\rm s.c.}^n \ge \overline{\Sigma}_{\rm in.disk}^n\,\, \mathrm{and} \,\, v_r(r_{\rm s.c.})\ge0 \,\, \mathrm{then} \nonumber\\
 \Sigma_{\rm s.c.}^{n+1}= \Sigma_{\rm s.c.}^n &-& \Delta M_{\rm flow}/S_{\rm s.c.} \nonumber\\
 M_\ast^{n+1}&=& M_\ast^n. \nonumber
\end{eqnarray}
Here, $\overline\Sigma_{\rm in.disk}$ is the averaged surface density of gas in the inner active disk (the averaging is usually done over
one au immediately adjacent to the sink cell), $S_{\rm s.c.}$ is the surface area of the sink cell, and $v_r(r_{\rm s.c.})$ is the radial component of velocity at the sink-disk interface. We note that $v_r(r_{\rm s.c.})<0$ when the gas flows from the active disk to the sink cell and $v_r(r_{\rm s.c.})>0$ in the opposite case. The superscripts $n$ and $n+1$ denote the current and the updated (next time step) quantities. The exact partition between $\Delta M_\ast$ and $\Delta M_{\rm s.c.}$ is usually set to 95\%:5\%, meaning that most of the mass lands directly on the star and only a small fraction is retained by the sink. This corresponds to fast mass transport through the sink. The effect of the $\Delta M_* : \Delta M_{\rm s.c.}$ partition on the disk evolution is studied in \cite{Vorobyov2019}. The calculated values of $\Sigma_{\rm
s.c.}^{n+1}$ are used at the next time step as the inner boundary values for the gas surface density. The radial velocity and internal energy at the inner boundary
are determined from the zero gradient condition, while the azimuthal velocity is extrapolated from the active disk to the sink cell assuming a Keplerian rotation.

The rate may be both negative, meaning the flow of mass from the disk to the sink, and positive, meaning the opposite flow from the sink to the disk. The mass transport rate through the sink-disk interface is also used to calculate the net mass of gas in the sink and in the star \citep[for detail see][]{Kadam2019}.

The known gas mass in the sink cell is further used at as the inner boundary values for the surface density in the disk. The radial velocity and internal energy at the inner boundary are determined from the zero gradient condition, while the azimuthal
velocity is extrapolated from the active disk to the sink
cell assuming a Keplerian rotation. These inflow-otuflow
boundary conditions enable a smooth transition of the
surface density and angular momentum between the inner active disk and the sink cell, preventing (or greatly
reducing) the formation of an artificial drop in the surface density near the inner boundary. Finally, we note that the outer boundary condition is set to a standard free outflow, allowing material to  flow out of the computational domain, but
not allowing any material to flow in.

\subsection{Solution procedure}

The continuity and momentum equations~(\ref{eq:mass}) and (\ref{eq:momentum}) and also the energy equations (\ref{eq:energy1}), (\ref{eq:energy2}), and (\ref{eq:energy3}), depending on the adopted cooling-heating scheme, are solved 
{ on the polar grid ($r, \phi$)} using the operator-split solution procedure similar in methodology to the ZEUS-2D code \citep{SN1992}.  The computational domain extends from the sink cell boundary at $r_{\rm sc}=5$~au to the initial cloud core radius at $r_{\rm out}$ (see Table~\ref{table1}). The star (once formed) is located at the coordinate origin and the stellar motion in response to the disk potential is not taken into account in this study.
The adopted resolution is $512\times 512$ grid cells, which on the logarithmically spaced grid corresponds to a spatial resolution of 0.1~au at a radial distance of 7 au and 1.0~au at 70~au.
To correctly simulate disk fragmentation, the local Jeans length must be resolved by at least four numerical cells \citep{Truelove1998}. In the thin-disk limit, the Jeans length can be expressed as \citep{Vorobyov2013}
\begin{equation}
R_{\rm J} = {c_{\rm s}^2 \over \pi G \Sigma},
\end{equation}
where $c_{\rm s}$ is the sound speed and $G$ is the gravitational constant.
Fragments usually condense out of the densest sections of spiral arms at a typical distance of 100~au and then either migrate inward or scatter outward. The typical surface densities and temperatures in spiral arms do not exceed 100~g~cm$^{-2}$ and 100~K. Adopting these values, the corresponding Jeans length is 
$R_{\rm J}\approx$ 20~AU. The numerical resolution at 100~au is 1.4~au, thus fulfilling the Truelove criterion.

The solution is
split in the transport and source steps. In the transport
step, the update of hydrodynamic quantities due to advection
is done using the third-order piecewise parabolic interpolation scheme of \citet{Colella1984}. In the source step, the update of hydrodynamic quantities
due to gravity, turbulent viscosity, cooling, and heating is performed.  The gravitational potential 
of the matter in the computational domain is found by solving for the Poisson integral \citep{BT87}
\begin{eqnarray}
\label{potential}
   \Phi(r,\phi) & =&  \\ \nonumber
   &-& G \int_{\rm r_{\rm sc}}^{r_{\rm out}} r^\prime dr^\prime
                   \int_0^{2\pi}
                \frac{\Sigma (r^\prime,\phi^\prime) d\phi^\prime}
                     {\sqrt{{r^\prime}^2 + r^2 - 2 r r^\prime
                        \cos(\phi^\prime - \phi) }}  \, ,
\end{eqnarray} 
{We note that we do not introduce an explicit smoothing length when calculating the integral~(\ref{potential}), as was advocated in \citet{Mueller2012}. This is because our method for calculating the integral already includes an implicit smoother set equal to the size of the grid cell in which the potential is calculated \citep[see eq. 2-206 in][]{BT87}. Since the size of the cell and the disk scale height are both linearly proportional to radial distance in our model, our implicit smoothing  length is also linearly proportional to the disk scale height, in agreement with  \citet{Mueller2012}, but the coefficient of proportionality may be different. }

We use an explicit integrator to compute the viscous force and heating (the last terms on the right-hand side of the momentum and internal energy equations). This is found to be adequate as long as the $\alpha$-parameter does not exceed greatly 0.01. The update of the internal energy per surface area in the $\beta$-cooling scheme is done using an analytic solution, while in ThES1 and ThES2 the update due to cooling and heating is done implicitly using the Newton-Raphson method of root finding, complemented by the bisection method where the Newton-Raphson iterations
fail to converge. The implicit solution is applied to avoid too small time steps that may emerge in regions of fast heating or cooling. A small amount of artificial viscosity is added to smooth out the shocks, which may occur in the gas flow, but the associated torques are much smaller than those due to turbulent viscosity. 


We solve non-equilibrium kinetic equations for H, H$_{2}$, H$^{+}$, D, HD, D$^{+}$, and e, while H$^{-}$ fraction is calculated from the equilibrium of 
reactions 3, 4, 11, 12, 15, and 16 in Table~\ref{Table:reactions}.
The method of calculating the rate coefficients of the reverse reactions using the rate coefficients of the forward reactions (summarized 
in Table~\ref{Table:reactions}) is explained in Appendix C in \cite{Matsukoba2019}. 
We assume that helium is always neutral and its fractional abundance is $y_{\mathrm{He}}=8.333\times10^{-2}$. 

 We further assume that our species are
collisionally coupled  with gas, which eliminates the need for solving separate equations of
motion for each species.  The remaining continuity equation for the surface density ($\Sigma_{i}$) of each of the species is written as
\begin{equation}
    \frac{\partial\Sigma_{i}}{\partial t}+\bl{\nabla}_{p}\cdot\left(\Sigma_{i} \bl{v}_{p}\right) = k_{j,k} \Sigma_j \Sigma_k - k_{k,i} \Sigma_k \Sigma_i,
    \label{eq:species}
\end{equation}
where  the right-hand terms are the sources and sinks due to
chemical reactions. The set of Equations~(\ref{eq:species}) is solved in two steps. First, $\Sigma_i$ are updated by solving implicitly the set of non-equilibrium kinetic equations taking chemical reactions into account. This step is performed between the source and transport steps of the hydrodynamic part. Then, 
the chemical species are  advected with the gas flow using the same third-order-accurate scheme of \citet{Colella1984}.

\begin{figure}
\begin{centering}
\includegraphics[width=1\columnwidth]{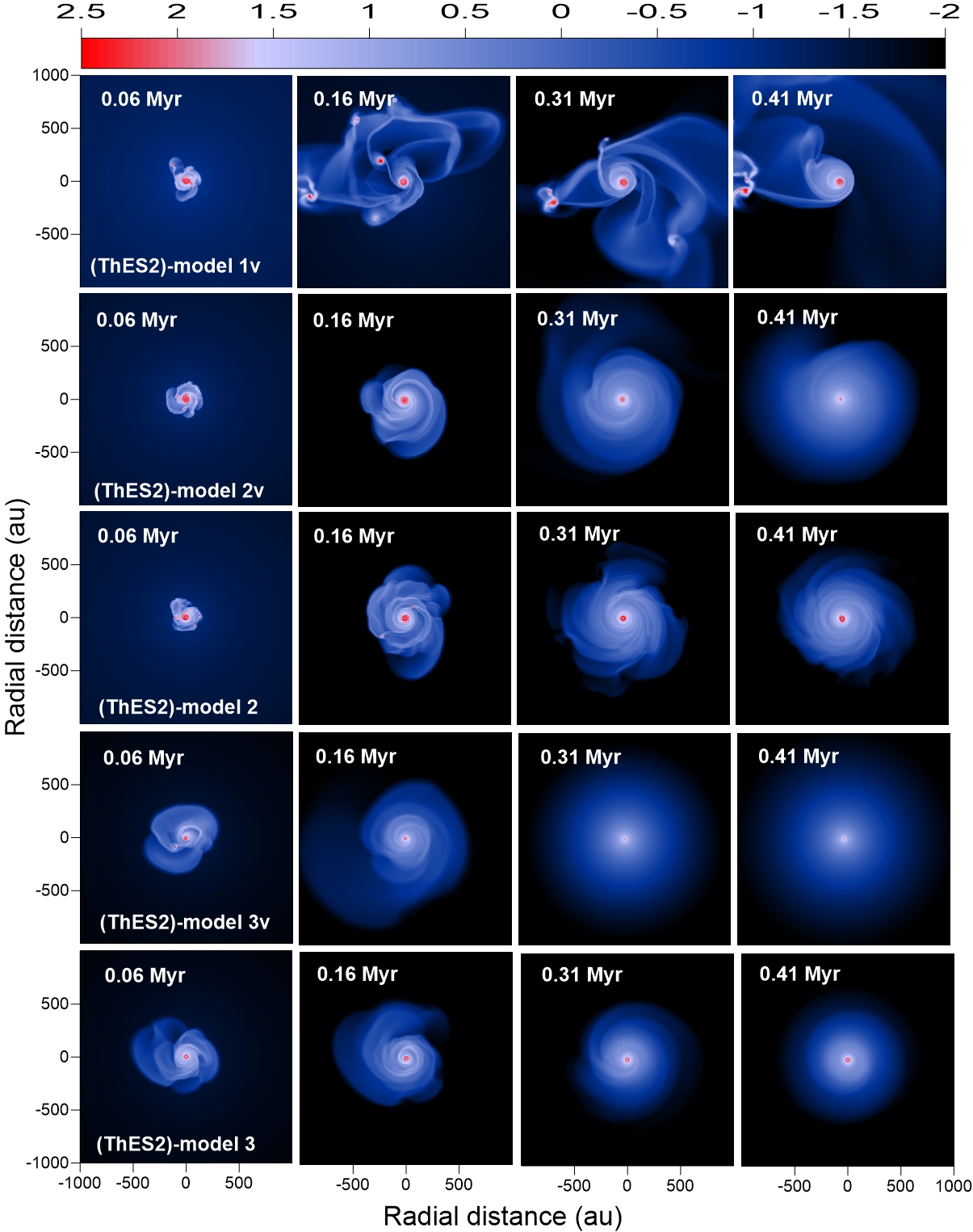}
\par\end{centering}
\caption{\label{fig:1} Gas surface density distributions in the five models considered. Each row presents a specific model as indicated and each column corresponds to a specific time starting from disk formation. The scale bar is in log g~cm$^{-2}$. }
\end{figure}

\begin{figure}
\begin{centering}
\includegraphics[width=1\columnwidth]{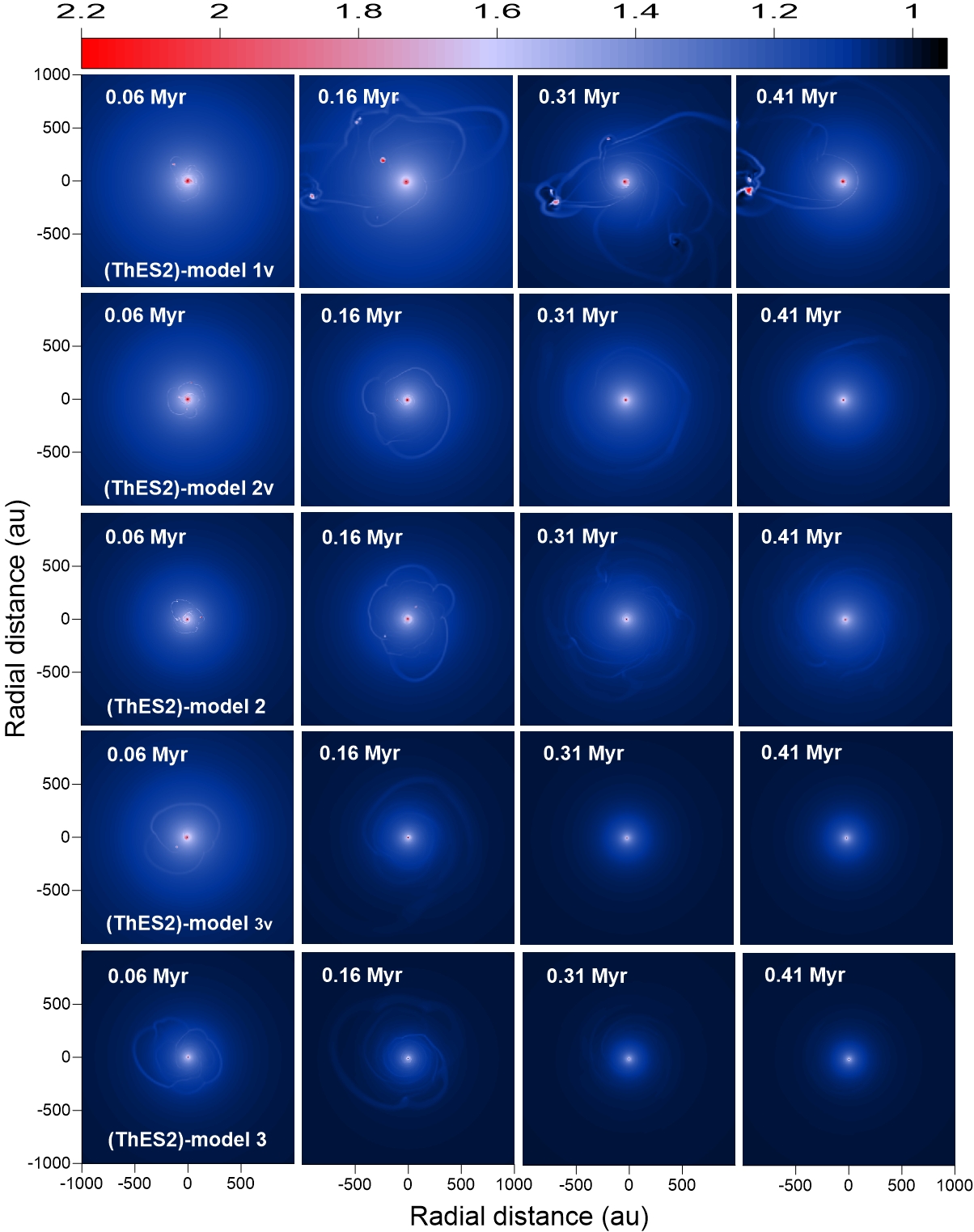}
\par\end{centering}
\caption{\label{fig:2} Similar to Figure~\ref{fig:1}, but for the dust temperature. The scale bar is in log K.}
\end{figure}

\section{Disk evolution in ThES2}
\label{newmodel}
We start with describing the disk evolution in the framework of the most elaborate thermal evolution scheme ThES2 with separate gas and dust temperatures. Figure~\ref{fig:1} presents the gas surface density distribution in the inner $2000 \times 2000$~au$^2$ box for the five considered models. The most massive model (in terms of the pre-stellar core) is shown in the top row, while the least massive model with different $\alpha$-values ($10^{-2}$ and $10^{-4}$) is shown in the two bottom rows. In between, the intermediate-mass model is shown, also for the two values of $\alpha$-parameter.

Clearly, the mass of the pre-stellar core determines the properties of the disk that form as a result of gravitational collapse. In the most massive (ThES2)-model~1v the disk is strongly fragmented during the considered evolution period (up to 0.41~Myr), while in the least massive (ThES2)-model~3v and (ThES2)-model~3 the disk shows signatures of fragmentation only in the very early stages ($< 0.1$~Myr) and becomes virtually axisymmetric in the later evolution. Low turbulent viscosity in (TheS2)-model2 and (ThES2)-model3 helps gravitational instability last longer, in agreement with the recent findings of \citet{Rice2018}. This is  because viscosity not only acts to smooth out local inhomogeneities, but also reduces the net disk mass due to an elevated mass transport.  The disks in the low-viscosity models are also more compact due to the lack of viscous spreading. An increased rate of pre-stellar core rotation, as indicated by a higher $\beta_{\rm c}$-value in (ThES2)-model~3v and (ThES2)-model~3, does not offset the effect of a decreased initial core mass. The latter two models show weaker and less persistent signatures of gravitational instability and fragmentation. Although higher $\beta_{\rm c}$ models can form more extended disks (gravitational instability and fragmentation are strongest at large distances), the higher $M_{\rm core}$ models form more massive disks and this factor appears to be decisive for the development and sustainability of gravitational instability and fragmentation.

\begin{figure}
\begin{centering}
\includegraphics[width=1\columnwidth]{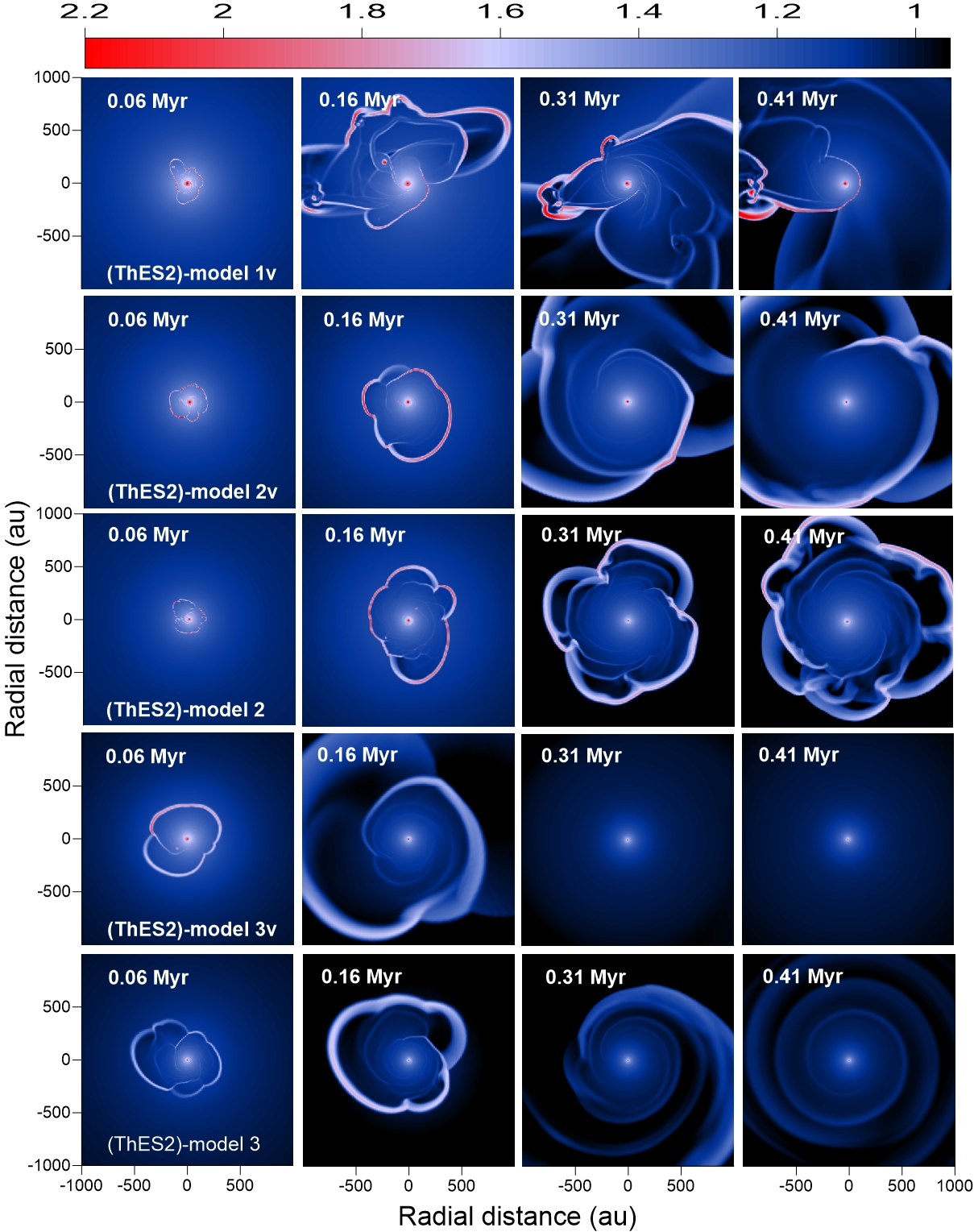}
\par\end{centering}
\caption{\label{fig:3} Similar to Figure~\ref{fig:1}, but for the gas temperature. The scale bar is in log K.}
\end{figure}

Figure~\ref{fig:2} presents the spatial distribution of dust temperature in the inner $2000\times2000$~au$^2$ for the five considered models.  Overall, the higher mass models are warmer than their lower-mass counterparts, which can be explained by  a higher stellar luminosity feedback in the models that form from more massive cores. Indeed, the terminal stellar masses in (ThES2)-model~1v and (ThES2)-model~2v are 0.67~$M_\odot$ and 0.43 $M_\odot$, respectively. Stars with such masses have photospheric luminosities of 3.3~$L_\odot$ and 1.8~$L_\odot$, according to the adopted stellar evolution tracks from \citet{Yorke2008}. The low-mass (ThES2)-model~3v has the terminal stellar mass of 0.18~$M_\odot$ and its photospheric luminosity is only 0.3~$L_\odot$. A similar trend is found for the accretion luminosities -- higher-mass models have higher accretion luminosities thanks to higher accretion rates driven by more massive (and more gravitationally unstable) disks. Viscous models also have higher disk temperatures, which can be explained by additional viscous heating. Finally, we note that the gaseous clumps formed via disk fragmentation are distinguished by higher dust temperatures as compared to the immediate disk environment. This temperature increase is caused by compressional heating (PdV work) of gravitationally bound and optically thick clumps. 


One interesting feature that can be noted in Figure~\ref{fig:2} is a slight temperature increase at the outer edge of the disk. This effect is however most pronounced  in Figure~\ref{fig:3}, which shows the gas temperature distribution in the inner $2000 \times 2000$ au$^2$ box for the five considered models.  Clearly, the gas temperature distribution is strongly non-monotonic -- the gas temperature generally declines with radius but there exists a high-temperature rim in the disk outer regions, where the gas temperature can exceed 100~K. We note that the gas temperature in the immediate surroundings is just a few tens of Kelvin. 

\begin{figure}
\begin{centering}
\includegraphics[width=1\columnwidth]{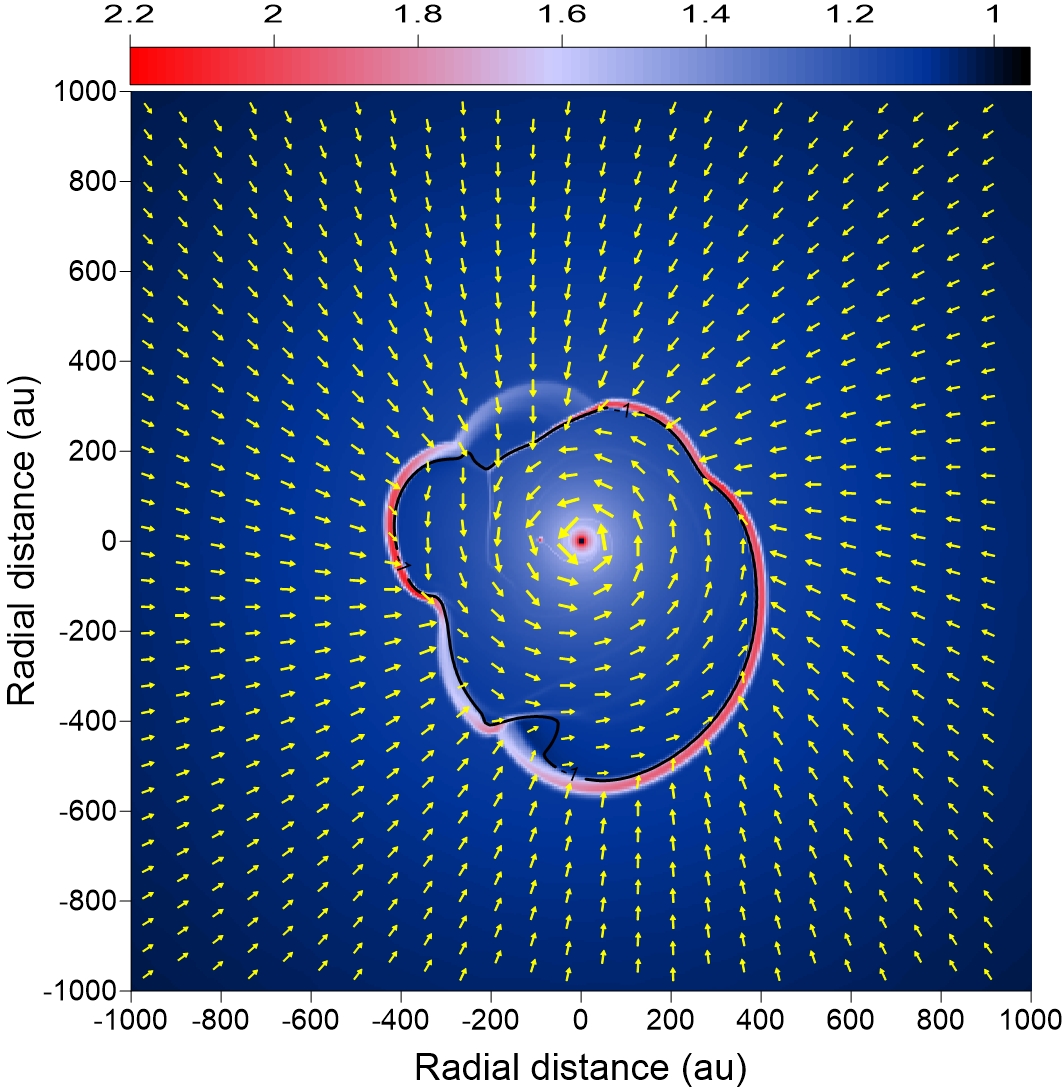}
\par\end{centering}
\caption{\label{fig:3a} Gas velocity field superimposed on the gas temperature distribution in (ThES2)-model~2v. The black contour line outlines a gas surface density of 0.1~g~cm$^{-2}$.}
\end{figure}

To better illustrate the origin of the jump in the gas temperature distribution, we plot in Figure~\ref{fig:3a} the gas velocity field superimposed on the gas surface density distribution in (ThES2)-model~2v at $t=0.16$~Myr. The black contour line defines the disk regions where the gas surface density is equal to 0.1~g~cm$^{-2}$, a value below which protoplanetary disks usually have a sharp outer edge \citep[see fig. 8 in][]{Andrews2009}. Clearly, the jump in the gas temperature occurs near the disk outer edge, where the gas surface density is low and where the infalling matter from the envelope meets the rotating disk. The converging gas flows produce additional compressional heating to the gas component, but the low surface densities of gas and dust prevent the gas from quickly attaining thermal equilibrium with the dust through mutual collisions. As a result, the gas temperature decouples from that of dust. This means that the gas temperature jump is expected to be most pronounced in the embedded stages of disk evolution, which seems to be the case in Figure~\ref{fig:3}.
The strength of the gas temperature jump diminishes with time in the intermediate- and low-mass models 2 and 3, for which the embedded phase ends at $t=0.19$~Myr and $t=0.13$~Myr, respectively (the end of the embedded phase is set to the time instance when less than 5\% of the initial pre-stellar core mass still resides in the infalling envelope). The high-mass model~1 remains in the embedded phase for the entire duration of our simulations.

\begin{figure}
\begin{centering}
\includegraphics[width=1\columnwidth]{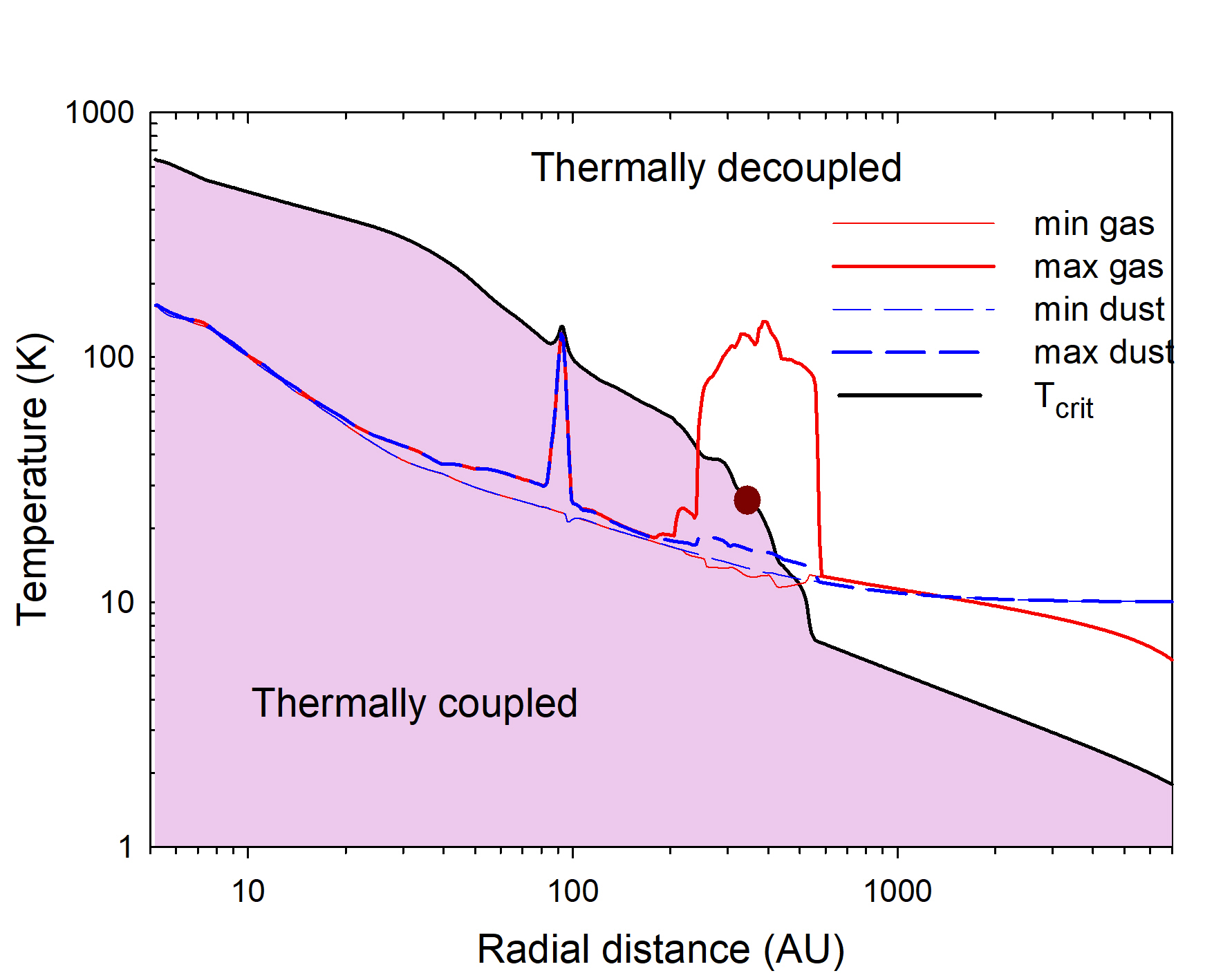}
\par\end{centering}
\caption{\label{fig:3c}  Decoupling of gas and dust temperatures in the disk and envelope in (ThES2)-model~2v at t=0.16~Myr. Shown are the threshold temperature ($T_{\rm crit}$, black thick line) above/below which the gas and dust temperatures are thermally decoupled/coupled. The red thin and thick lines present the minimum and maximum azimuthal variations of the gas temperature, respectively, while the blue dashed thin and thick lines show the corresponding quantities for the dust temperature. The brown circle marks the position of the disk outer edge. }
\end{figure}

The reason for decoupled gas and dust temperatures can be understood from the following analysis. Assuming that radiative cooling of dust is balanced by collisional heating with gas, Equation~(\ref{Eq:dust_balance}) can be expressed as follows
\begin{equation}
    T_{\rm d}\simeq 120~\mathrm{K} \left( {T_{\rm g} \over 100 \mathrm{K}}~\right)^{0.3} \left( {n_{\rm g} \over 10^{10}~\mathrm{cm}^{-3}} \right)^{0.2},
\end{equation}
where $n_{\rm g}$ is the number density of gas. By setting $T_{\rm g} = T_{\rm d}$ we define the threshold temperature above which gas and dust thermally decouple from each other. This threshold temperature can be written as 
\begin{equation}
T_{\rm crit} \simeq 130~\mathrm{K} \left( {n_{\rm g} \over 10^{10}~\mathrm{cm}^{-3}} \right)^{0.3}. 
\end{equation}

To illustrate the effect of threshold temperature, we take 
(ThES2)-model~2v at $t=0.16$~Myr and plot $T_{\rm crit}$ as  
a function of radial distance by the thick black line in Figure~\ref{fig:3c}. We used the azimuthally averaged gas number density when calculating $T_{\rm crit}$.   The red thin and thick solid lines show the minimum and maximum azimuthal variations in the gas temperature, respectively, while the red thin and thick dashed lines present the corresponding variations for the dust temperature. Note that gas and dust temperatures coincide inside 200~au where both temperatures are lower than the threshold one. This is a thermally coupled region of the disk. The variations in gas and dust temperatures begin to deviate from each other beyond 200~au. In particular, the gas temperature becomes systematically higher than the threshold temperature $T_{\rm crit}$, meaning that the disk is now in a thermally decoupled state.

\begin{figure}
\begin{centering}
\includegraphics[width=1\columnwidth]{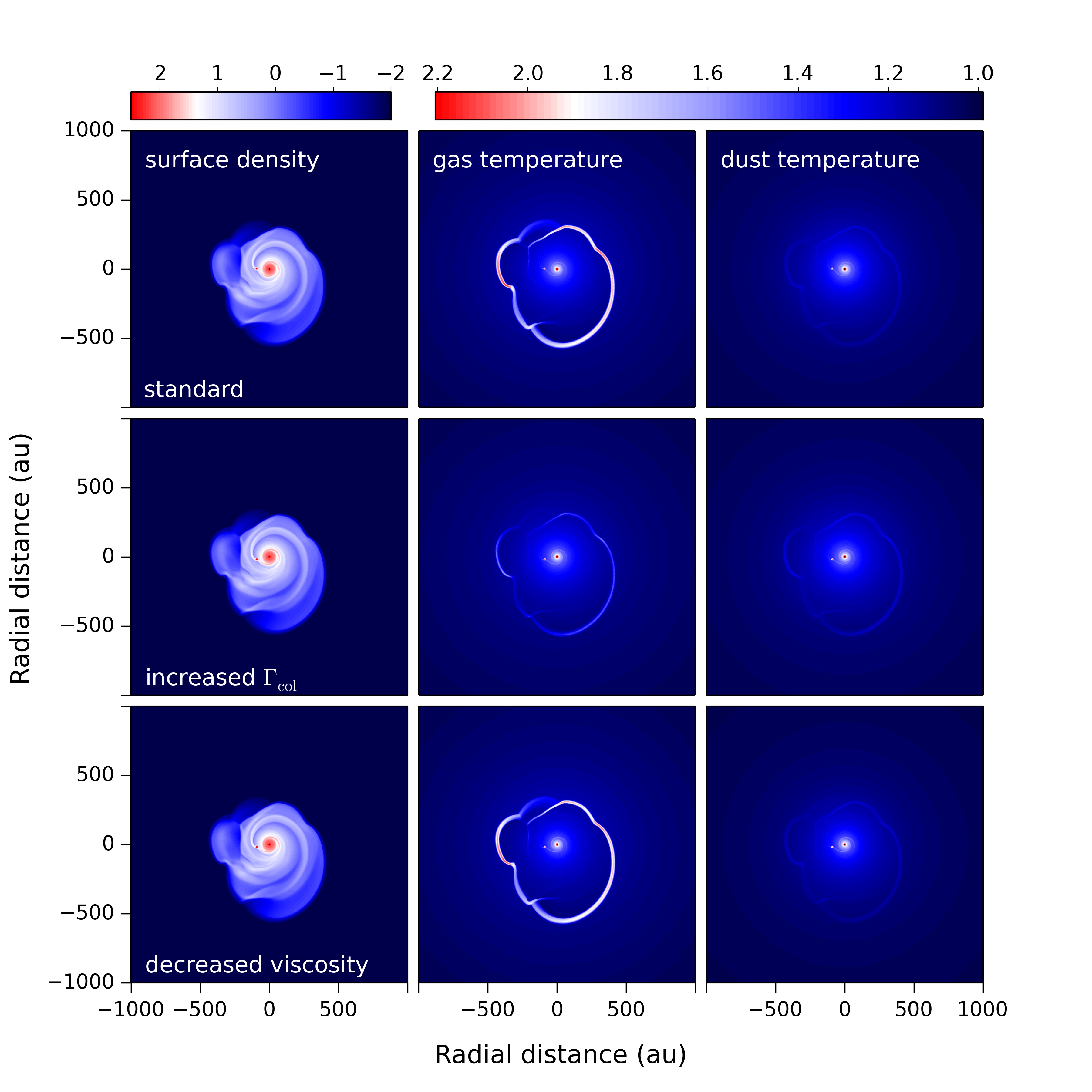}
\par\end{centering}
\caption{\label{fig:3b} Comparison of the gas surface density, gas and dust temperatures in model~2. The top and bottom rows correspond to (ThES2)-model~2v and (ThES2)-model~2, while the middle row presents results for a test model with the rate of collisional heat exchange between dust and gas $\Gamma_{\rm coll}$ increased artificially by a factor of 50. }
\end{figure}

To elaborate further on the cause for the gas temperature jump, we considered a test model in which we artificially increased the rate of collisional energy exchange between dust and gas ($\Gamma_{\rm coll}$) by a factor of 50. This exercise mimics an increase in the density of both material species without affecting the integrity of the disk. If the temperature jump is due to slow exchange of energy between compressionally heated gas and radiatively cooled dust, then the jump should diminish as we increase $\Gamma_{\rm coll}$. Figure~\ref{fig:3b} demonstrates that this is indeed the case.  The top row presents the gas surface density, gas and dust temperatures for the standard (ThES2)-model~2v at t=0.16~Myr. The middle row shows the same quantities in a test model with $\Gamma_{\rm coll}$ increased artificially by a factor of 50. Clearly, the gas and dust temperatures in this test case are similar and the gas temperature jump near the outer disk edge is greatly reduced. The bottom row presents the resulting distributions for (ThES2)-model~2 with a reduced rate of viscous heating. As can be seen, the rate of viscous heating does not affect notably the strength of the gas temperature jump.
We conclude that gas-to-dust energy exchange defined by Equation~(\ref{Eq:ColHeat}) is the most important mechanism to capture the effect of temperature decoupling. It sets the dust temperature through Equation~(\ref{Eq:dust_balance}), and the resulting dust temperature enters the dominant $Q_{\rm cont}$ term in Equation~(\ref{CoolingAll}). The second in importance is $Q_{\rm metal}$ term, but its effect is notable only near the disk outer edge. We note that ThES2 can be applied to a wide range of metallicities and at lower metallicities other terms in Equation~(\ref{CoolingAll}) can become important.

\begin{figure*}
\begin{centering}
\includegraphics[width=2\columnwidth]{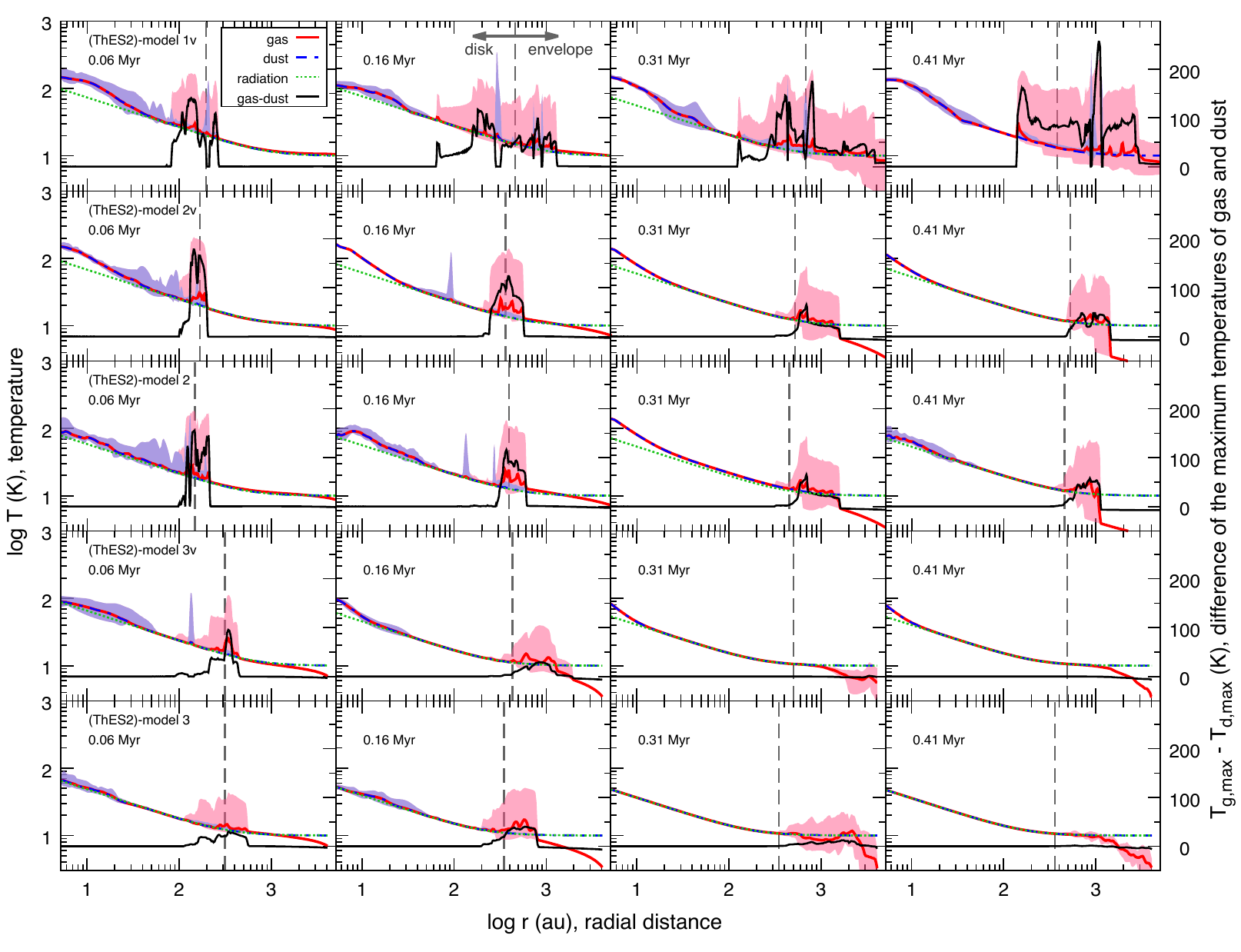}
\par\end{centering}
\caption{\label{fig:5} Azimuthally averaged radial distributions of gas (red solid line), dust (blue dashed-dotted line), and radiation (green dotted line) temperatures in five models considered. The shaded areas indicate the range of azimuthal variations of gas (pink) and dust (blue) temperatures at each radius. The black line is the difference between the maximum temperatures of gas and dust (see the right-hand-side axis). The gray dashed line shows the radius of the disk outer edge. The arrangement of panels is the same as in Figure 1.}
\end{figure*}

What could be the consequences of decoupling between the gas and dust temperatures? We will see later in Sect.~\ref{oldmodel} that this decoupling does not have a notable effect on the disk structure and propensity to gravitational instability and fragmentation. However, an increase in the gas temperature near the disk outer edge may have important consequences for the chemical processing of gas that flows in from the envelope. An increase in the gas temperature to more than a hundred Kelvin could launch gas phase reactions that are expected to be  dormant in these otherwise cold disk outer regions. For instance, \citet{Oya2016} inferred a local increase in the gas kinetic temperature in the disk outer regions of IRAS~16293-2422 based on the peculiar chemical composition and explained this feature by a possible shock heating at the disk-envelope interface. The observational detection of a gas temperature jump is however not unambiguous, as more recent observations of IRAS~16293-2422 revealed no such structures \citep{Hoff2020}. Our numerical simulations also suggest that these features are not omnipresent and their occurrence depends on the disk evolution stage.

Decoupling of gas and dust temperatures may also affect the growth rate of small (sub)micron-sized dust particles which flow in with gas from the envelope. If volatile species become oversaturated in the warm gas environment near the disk outer edge, this may facilitate the growth of icy mantles on cold dust particles (a similar effect can be observed in a Turkish bath when water vapor condenses on cold objects that are brought to the bath).   

We note that the temperature decoupling between gas and dust may not only be limited to the outer disk regions.
Recent studies have already demonstrated the importance of radiative disk properties on the formation and position of gaps, spirals, and snow lines in protoplanetary disks 
{ \citep{Zhang2020,Ziampras2020}.} The formation of gaps and rings in the dust density distribution can also lead to a reduced rate of energy transfer between gas and dust in the regions of depressed density (i.e., gaps), possibly resulting in temperature decoupling. This may have important consequences for the disk mass estimates which sensitively depend on the assumed disk temperature. We plan to explore this effect in follow-up studies.


Finally, in Figure~\ref{fig:5} we make a detailed comparison of the azimuthally averaged radial gas, dust, and irradiation temperature profiles in all five models considered. In particular, the red and blue curves present the gas and dust temperatures, while the green dotted curve is the temperature of stellar and background irradiation. Let us first consider the black line which illustrates the maximum deviation of the gas temperature from that of dust (see the right-hand axes).  Clearly, this deviation can reach hundreds of Kelvin in the massive and intermediate-mass models 1 and 2, especially in the early stages of disk evolution. The deviation peaks in the  the disk outer regions in the vicinity of the disk outer edge (marked with the vertical dashed lines) and diminishes in the inner parts of the disk. 

We also note that the gas and dust temperatures show considerable azimuthal variations as illustrated by the shaded areas: tinted with blue for the dust temperature variations and with pink for the gas temperature variations. These variations reflect the underlying non-axisymmetric distribution of gas in the disk and the circumdisk environment (see Fig.~\ref{fig:1}). In the inner disk regions variations in both temperatures are similar (that is why only the blue shaded area is visible). In the outer disk regions, however, the variation in the gas temperature greatly exceeds that of dust. Interestingly, the gas temperature in the regions beyond 1000 au drops systematically below that of dust (which is 10~K, set by the background irradiation). This is an inverse effect compared to that found for the disk outer edge where the gas temperature exceeds that of dust. In the regions beyond 1000 au the only notable heating mechanism is the background irradiation, which directly sets the dust temperature. Extremely low gas densities, however, prevent dust and gas temperatures from equalizing through mutual collisions, leading to progressive decoupling between the two temperatures.


\section{Comparison of disk evolution in ThES1 and ThES2}
\label{oldmodel}

\begin{figure*}
\begin{centering}
\includegraphics[width=2\columnwidth]{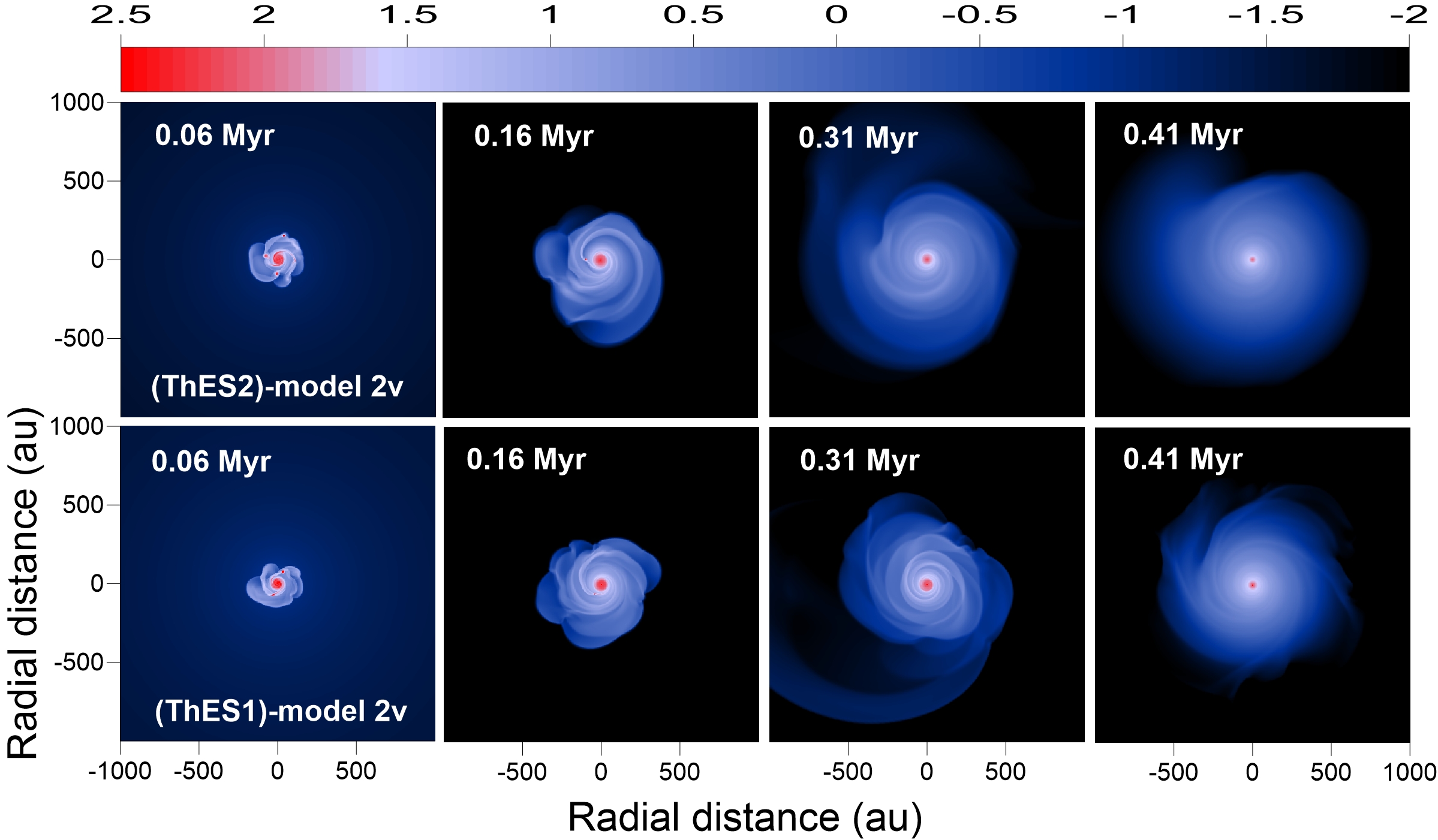}
\par\end{centering}
\caption{\label{fig:6} Comparison of gas surface density distributions in model~2v with ThES2 (separate dust and gas temperatures) and ThES1 (equal gas and dust temperatures). The scale bar is in log g~cm$^{-2}$. }
\end{figure*}

In this section, we compare the disk evolution for two different thermal schemes ThES1 and ThES2. Our motivation is to find out if the disk evolution with separate dust and gas temperatures (ThES2) can be notably different from the disk evolution with equal dust and gas temperatures (ThES1). For this purpose, we have chosen model~2v with the $\alpha$-value set equal to $10^{-2}$. Figure~\ref{fig:6} presents the gas surface density distributions in the inner $2000 \times 2000$ au$^{2}$ box for (ThES2)-model~2v (top row) and (ThES1)-model~v2 (bottom row).  Somewhat surprisingly, the overall evolution is similar whether we consider ThES2  or ThES1.  The disks in both models are gravitationally unstable and are prone to fragmentation in the initial stages of evolution.

\begin{figure}
\begin{centering}
\includegraphics[width=1\columnwidth]{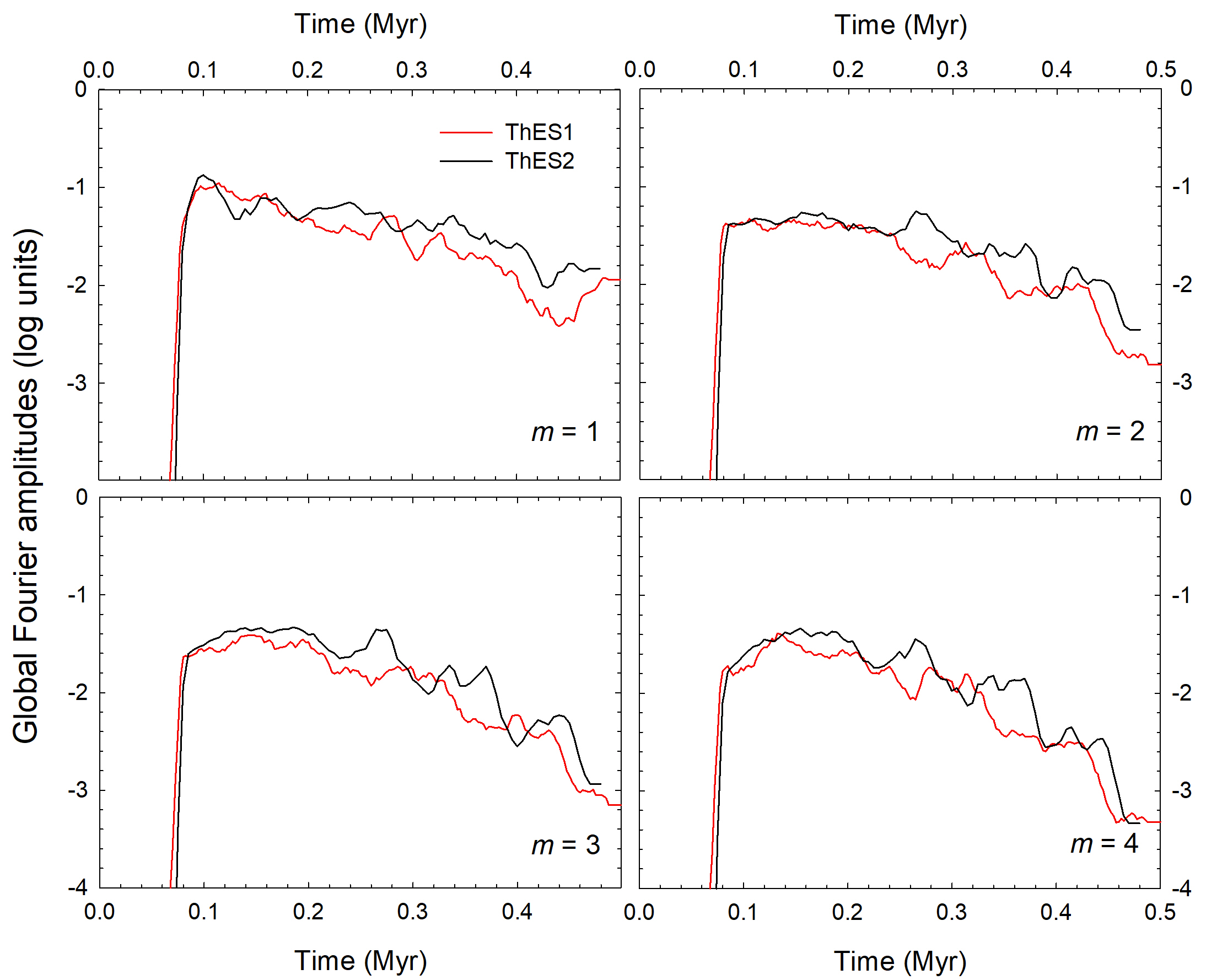}
\par\end{centering}
\caption{\label{fig:6a} Comparison of the global Fourier amplitudes in model~2v with ThES1 (red lines) and ThES2 (black lines). The global amplitudes for four modes ($m$=1,2,3, and 4) are shown in the four panels. The time is counted from the beginning of the core collapse. }
\end{figure}

To make a more quantitative analysis, we estimated the strength of gravitational instability by calculating the global Fourier amplitudes defined as 
\begin{equation}
C_{\rm m} (t) = {1 \over M_{\rm d}} \left| \int_0^{2 \pi} 
\int_{r_{\rm sc}}^{R_{\rm d}} 
\Sigma(r,\phi,t) \, e^{im\phi} r \, dr\,  d\phi \right|,
\label{fourier}
\end{equation}
where $M_{\rm d}$ is the disk mass, $R_{\rm d}$ is the disk's outer radius set for simplicity to 500 au, and $m$ is the ordinal number of the spiral mode.  When the disk surface density is axisymmetric, the amplitudes of all modes  that are not equal to zero vanish. When, say, $C_{\rm m}(t)=0.1$, the perturbation amplitude of  spiral density waves in the disk is 10\% that of  the underlying axisymmetric density distribution. The resulting Fourier amplitudes for (ThES1)-model~2v and (ThES2)-model~2v are shown in Figure~\ref{fig:6a}. It appears that (ThES2)-model~2v with distinct dust and gas temperatures is slightly more gravitationally unstable, but the difference in the Fourier amplitudes is insignificant. 

\begin{figure}
\begin{centering}
\includegraphics[width=1\columnwidth]{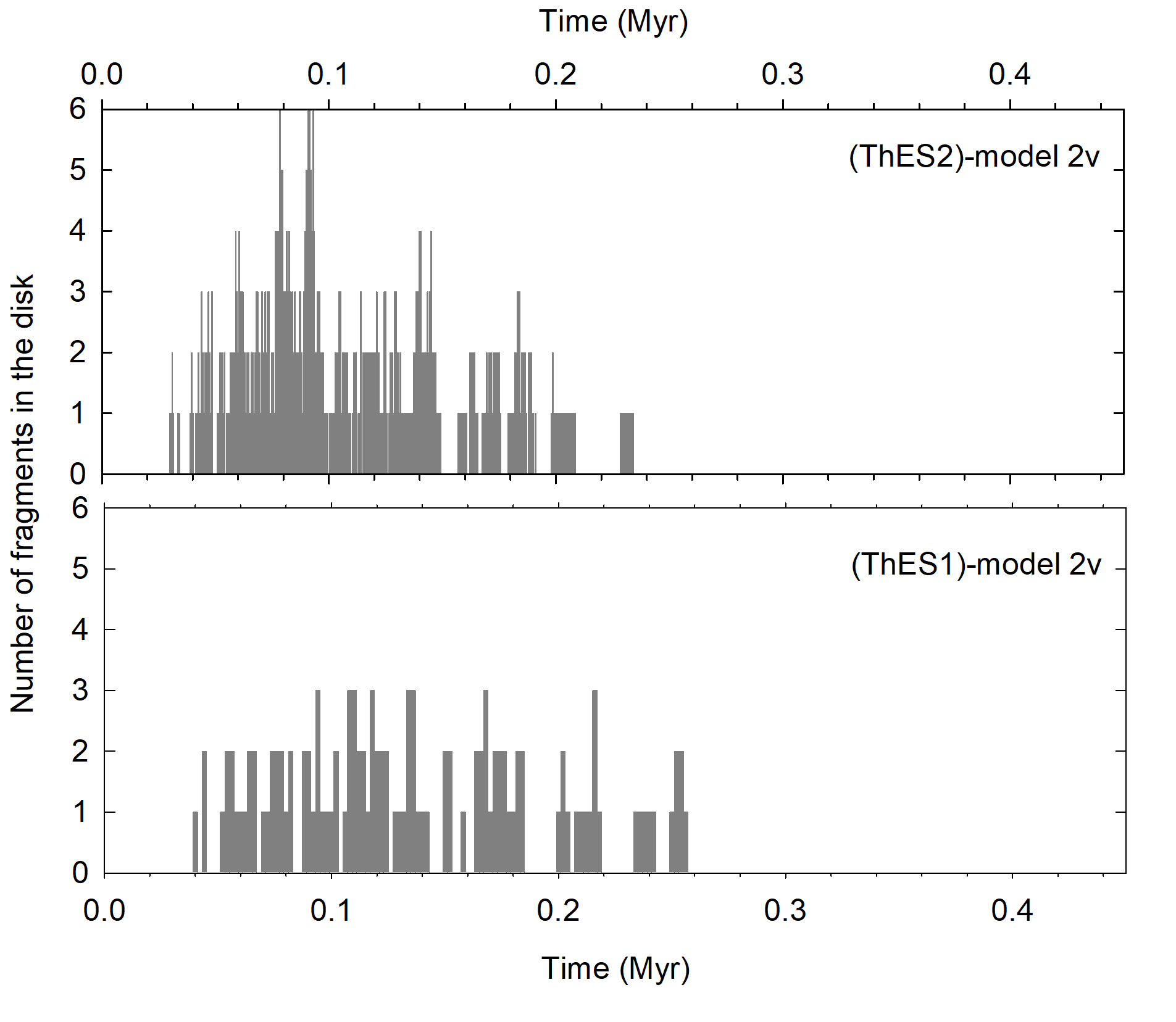}
\par\end{centering}
\caption{\label{fig:6b} Comparison of the number of fragments in the disk in model~2v with ThES2 (top panel) and ThES1 (bottom panel). The time is counted from the instance of disk formation. }
\end{figure}

Furthermore, we calculated the number of fragments in the disk at a given time instance formed through gravitational fragmentation using the fragment-tracking algorithm described in \citet{Vorobyov2013}. The results are presented in Figure~\ref{fig:6b} for (ThES2)-model~2v (top panel) and (ThES1)-model~2v. An increase in the number of fragments shows recent fragmentation, and a decrease shows either recent tidal destruction or accretion of the fragments on the star.
Again, the model with distinct dust and gas temperatures appears to form more fragments, but the model with equal dust and gas temperatures appears to sustain disk fragmentation for a longer time. To summarize, there are slight quantitative differences in the disk evolution for different thermal evolution schemes, but they do not cause qualitative changes in disk dynamics, such as suppression of disk fragmentation or disk stabilization against gravitational instability.


\begin{figure}
\begin{centering}
\includegraphics[width=1\columnwidth]{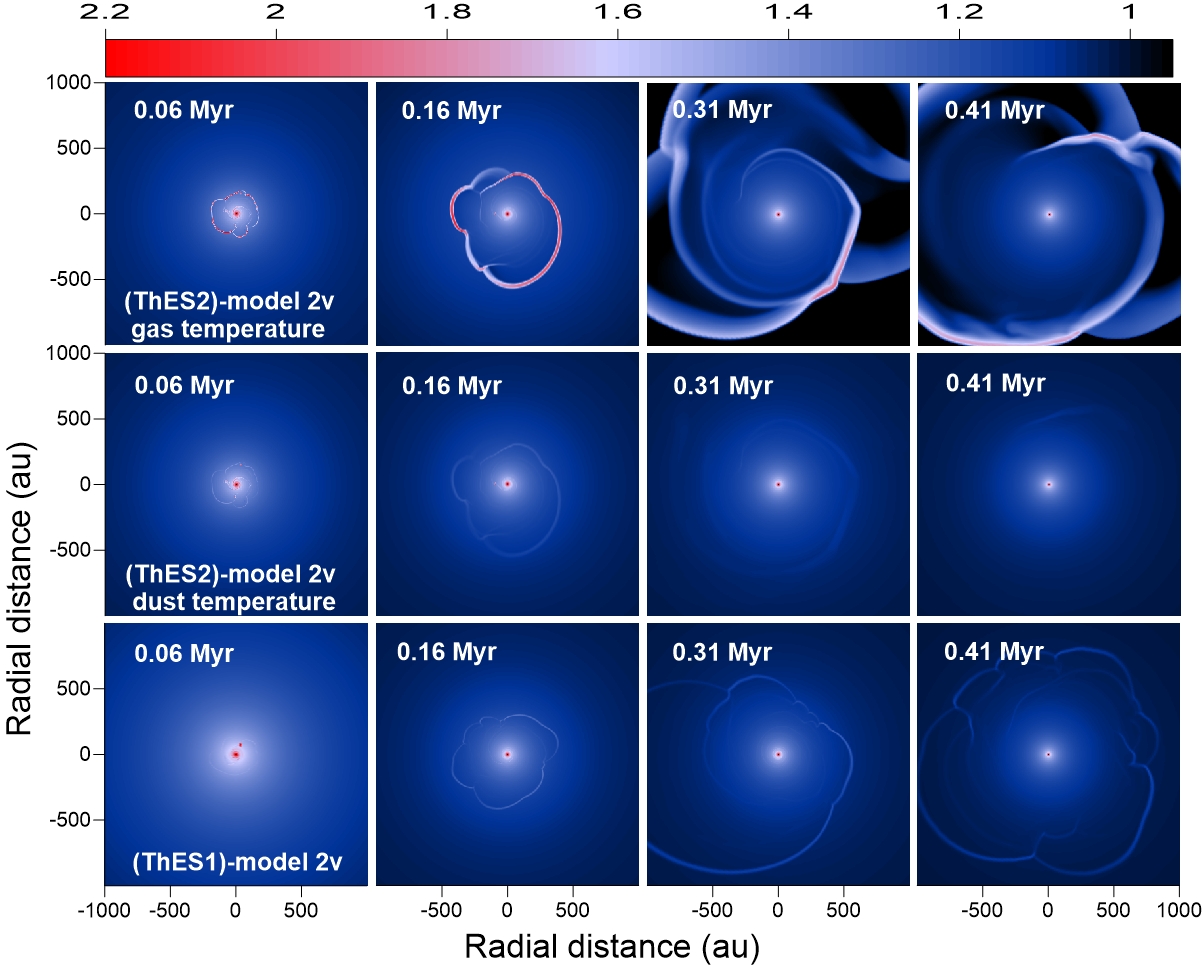}
\par\end{centering}
\caption{\label{fig:7} Comparison of gas and dust temperature distributions in model~v2 with ThES1 and ThES2. The first and second rows present the gas and dust temperature distributions in ThES2. The third row shos the temperature distribution (same for gas and dust) in ThES1. The bottom row provides a comparison of the azimuthally averaged temperatures in the two considered thermal evolution schemes. In particular, the solid lines show the temperatures of gas and dust, while the dashed lines provide the temperatures of stellar irradiation. The scale bar is in log Kelvin.}
\end{figure}

\begin{figure}
\begin{centering}
\includegraphics[width=1\columnwidth]{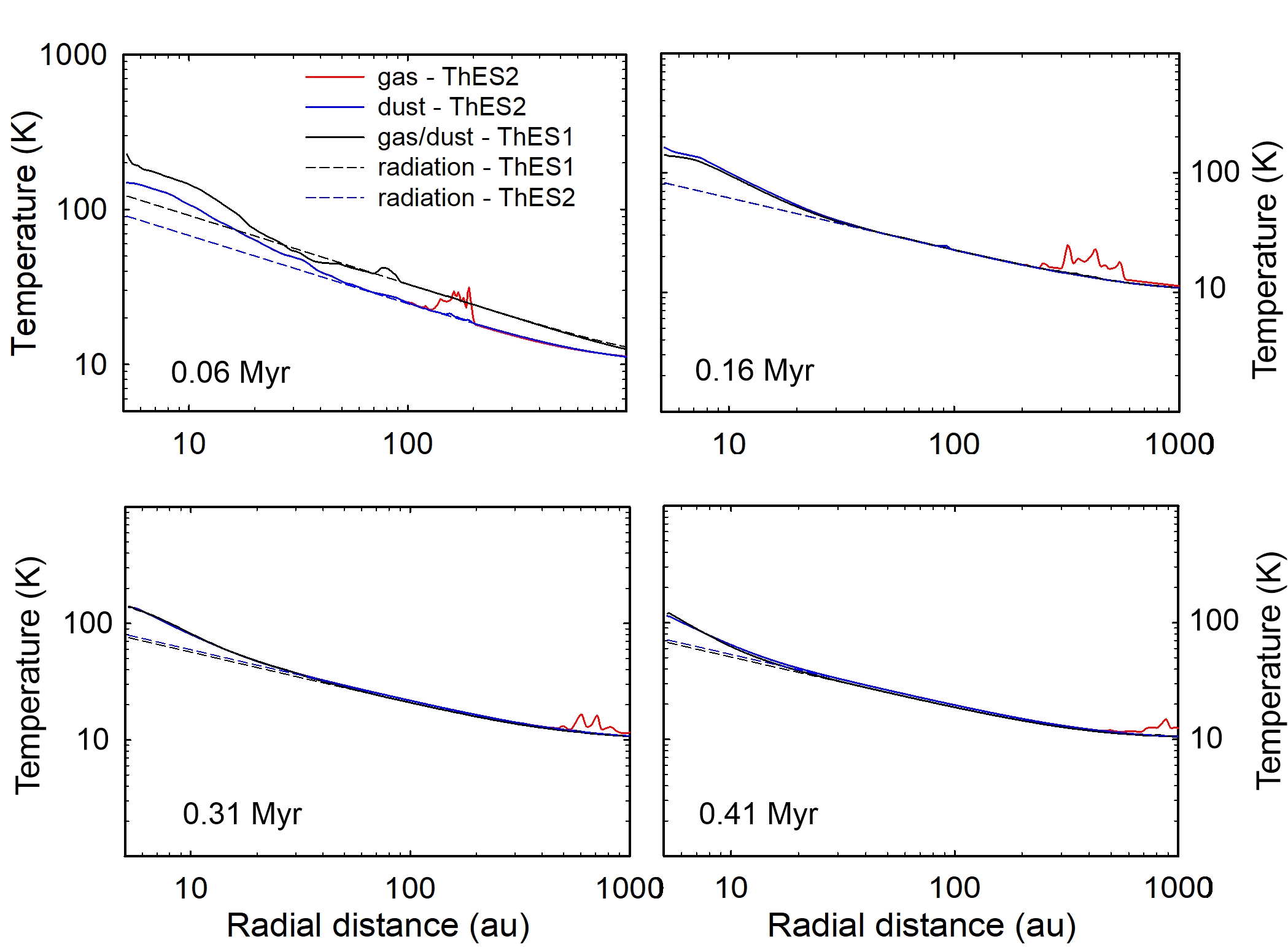}
\par\end{centering}
\caption{\label{fig:7a} Comparison of the azimuthally averaged gas and dust temperature distributions in model~v2 with ThES1 and ThES2. In particular, the solid lines show the temperatures of gas and dust, while the dashed lines provide the temperatures of stellar irradiation. Note that the gas and dust temperatures in the (ThES2)-model~v2 coincide everywhere except for the outer regions.}
\end{figure}

To further explore the difference in the evolution of model~2v with distinct thermal evolution schemes, we show in Figure~\ref{fig:7} the gas and dust spatial temperature distributions. More specifically, the first and second rows present the gas and dust temperature distributions in (ThES2)-model~2v, respectively, while the third row presents the temperature distribution (same for gas and dust) in (ThES1)-model~2v. { In addition, in Figure~\ref{fig:7a} we compare the azimuthally averaged gas and dust temperatures in the two considered models at the same evolutionary times as in Figure~\ref{fig:7}}. The solid lines show the gas and dust temperatures for different thermal schemes, while the dashed lines show the temperature of stellar and background irradiation. 

In the very early disk evolution, the gas and dust temperatures in (TheS1)-model~2v are systematically higher than in (ThES2)-model~v2, but this difference vanishes in the later evolution. Indeed, the azimuthally averaged gas and dust temperature profiles become almost indistinguishable, except for a well-defined jump in the gas temperature in the outer disk regions of (ThES2)-model~2v (separate gas and dust temperatures).  There also exists a notable positive deviation of the gas/dust temperatures from that of stellar irradiation in the inner disk regions, which occurs due to additional viscous heating of the disk.  The gas and dust temperatures are similar in the bulk of the disk because the collisional exchange of energy between gas and dust is efficient in equalizing the corresponding temperatures. Only in the outer disk regions this trend is broken thanks to decreased gas and dust densities and increased rates of compressional heating near the disk outer edge where the inflowing envelope lands at the disk (see Figures~\ref{fig:3a} and \ref{fig:3c}).
As the compressed gas moves closer to the star, it quickly cools through increased collisions with dust and dust radiative cooling.
The spatially limited extent of the disk regions where gas and dust temperatures decouple from each other may explain why the disk evolution weakly depends on the considered thermal evolution schemes.



\begin{figure}
\begin{centering}
\includegraphics[width=1\columnwidth]{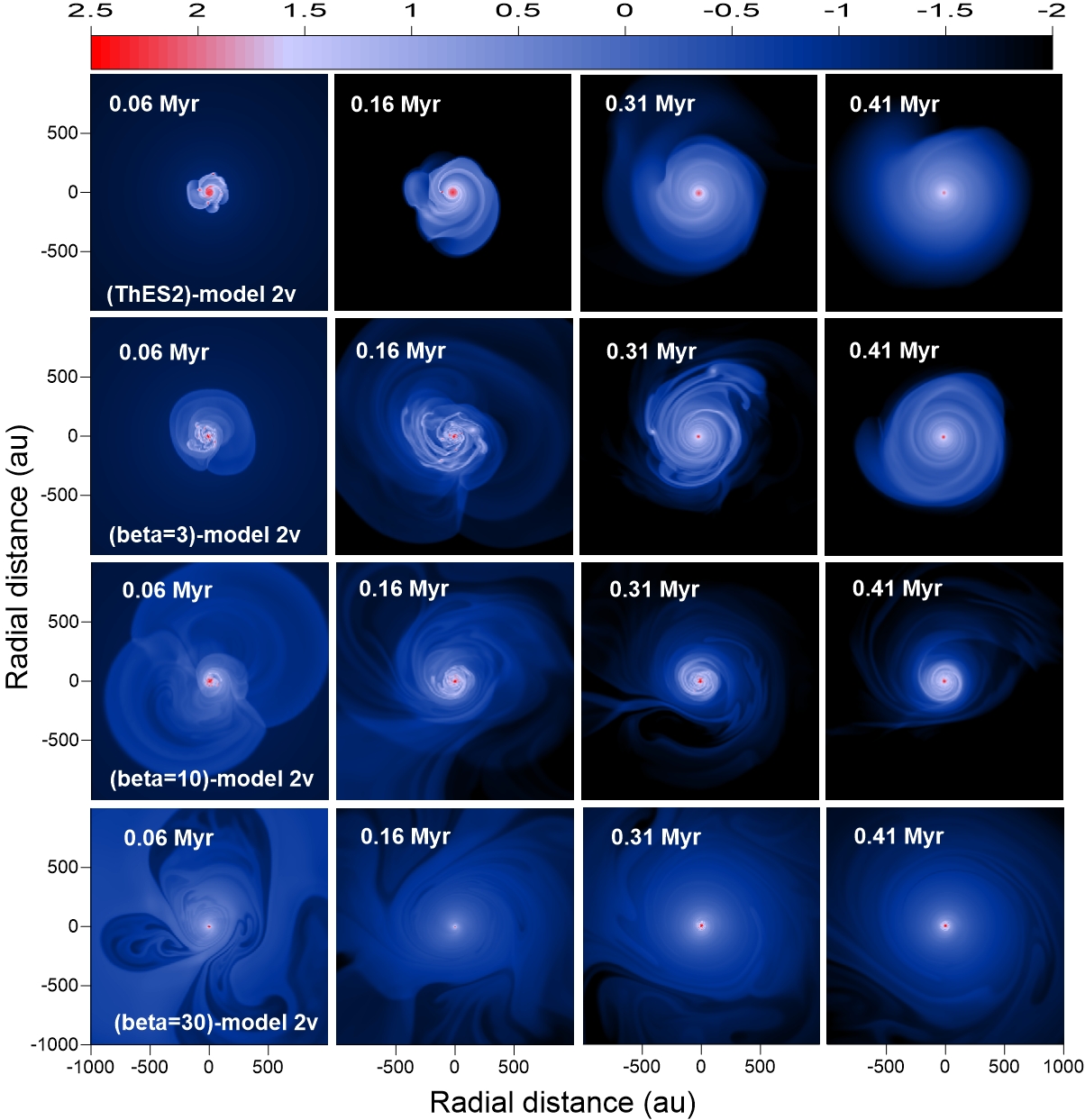}
\par\end{centering}
\caption{\label{fig:8} Comparison of gas surface density distributions in model~v2 with ThES2 and $\beta$-cooling. In particular, the first row corresponds to (ThES2)-model 2v, second row -- to (beta=3)-model~2v, third row -- to (beta=10)-model~2v, and bottom row -- to (beta=30)-model~2v. The time is counted from the instance of disk formation. The scale bar is in log g~cm$^{-2}$.}
\end{figure}

\section{Comparison with the $\beta$-cooling scheme}
\label{BetaCooling}
In this section, we compare the disk evolution in model~2v using two, opposite in its complexity, thermal evolution schemes: the most sophisticated ThES2 and the most simplified $\beta$-cooling. Our purpose is to determine if $\beta$-cooling can be used as a valid substitution for the more sophisticated thermal evolution schemes. We considered several values of the $\beta$-parameter and distinguish between the $\beta$-models by adding a prefix $beta$ to the model. For instance, (beta=3)-model~2v would correspond to model~2v with the $\beta$-cooling  scheme and $\beta$-value equal to 3.0.  In addition, we distinguish between the $\beta$-models with stellar and external irradiation by adding the suffix "Ir" to the $\beta$-value as in (beta=3Ir)-model~2v. We start with considering the case without irradiation and continue with the $\beta$-models taking irradiation into account.

\subsection{The case without irradiation}
Figure~\ref{fig:8} presents the gas surface density distributions in the inner $2000\times2000$~au$^{2}$ region for (ThES2)-model~2v (top row) and three models~2v with different $\beta$-values: (beta=3)-model~2v (second row), (beta=10)-model~2v (third row), and (beta=30)-model~2v (bottom row).  Clearly, the disk evolution in model~2v with ThES2 is notably different from that obtained with the $\beta$-cooling scheme. The $\beta=3$ model carries some resemblance to the model with ThES2, but is evidently more prone to gravitational fragmentation. The other two models with $\beta=10$ and $\beta=30$ are conspicuously different from the model with ThES2. The disk in these $\beta$-cooling models is much more extended initially and has a flocculent structure which is not typical of circumstellar disks.

\begin{figure}
\begin{centering}
\includegraphics[width=1\columnwidth]{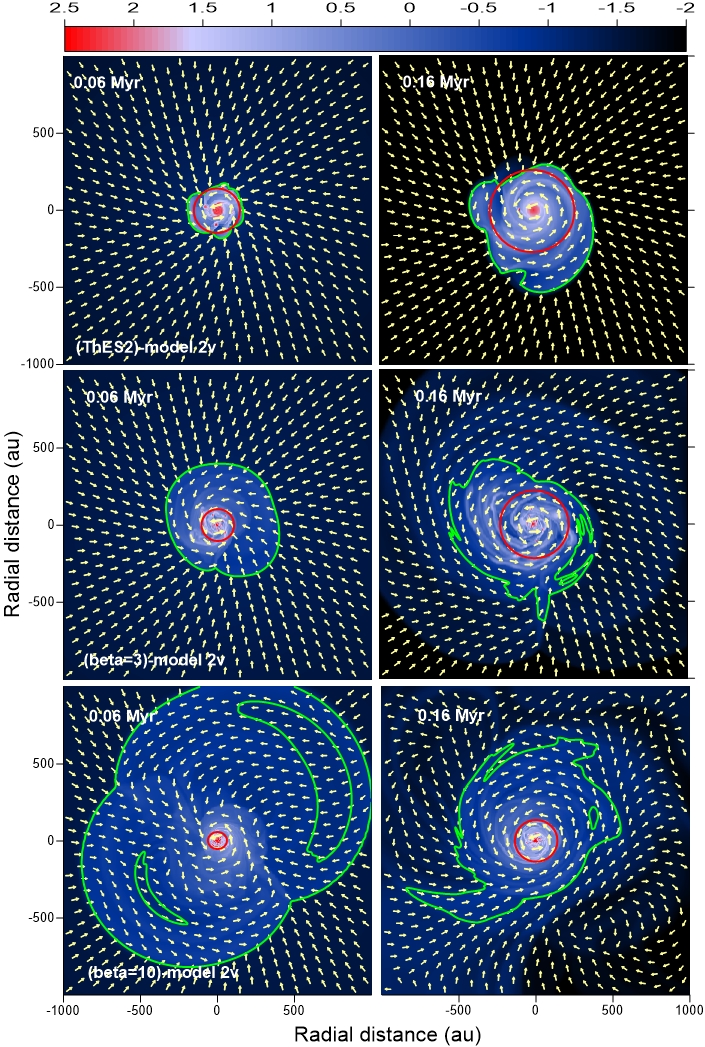}
\par\end{centering}
\caption{\label{fig:9} Gas velocity field superimposed on the gas surface density distribitions in (ThES2)-model~2v (top row), (beta=3)-model~2v (middle row), and (beta=10)-model~2v (bottom row at two evolutionary times as indicated in each panel. The red circles outline the radial distance beyond which the azimuthally averaged angular velocity deviates from the Keplerian rotation by more than 10\%. The green contour lines outline the regions where the gas surface density drops to 0.1~g~cm$^{-1}$. The time is counted from the instance of disk formation. The scale bar is in log g~cm$^{-2}$.}
\end{figure}

In fact, the high-density circumstellar structure in models with $\beta$-cooling is not a true centrifugally balanced disk with a near-Keplerian rotation, but rather a pseudo-disk with a significant deviation from circular motion. Figure~\ref{fig:9} presents the gas velocity fields superimposed on the gas surface density distributions in (ThES2)-model~2v (top row), (beta=3)-model~2v (middle row), (beta=10)-model~2v (bottom row) in the early stages of evolution. The red circles outline the disk regions within which the relative deviation of the azimuthally averaged angular velocity $v_\phi$ from the Keplerion rotation is smaller than 10\% (in fact, in most of this inner region it does not exceed 1-2\%). When calculating the Keplerian velocity we also took into account the contribution from the enclosed gaseous and dusty material. The green contour lines outline the radial extent beyond which the gas surface density drops below 0.1~g~cm$^{-2}$.

In the case of ThES2, the radial extent within which the gas surface density is greater than $0.1$~g~cm$^{-2}$ agrees rather well with the regions within which the deviation from the Keplerian rotation is smaller than 10\%. Some mismatch is seen at $t$=0.16~kyr to the south, but this is caused by a pronounced lopsidedness of the disk at this time instance. This means that $\Sigma=0.1$~g~cm$^{-2}$ may be regarded as the disk outer edge, as was already noted for protoplanetary disks in Ophiuchus by \citet{Andrews2009}. 
When we turn to the $\beta$-cooling models, however, the mismatch between the dense regions with $\Sigma>0.1$~g~cm$^{-2}$  and regions with near-Keplerian rotation becomes much more pronounced. This means that the dense structure outlined by the green contour lines is in fact a pseudo-disk with a significant (tens of per cent) deviation from the Keplerian motion.

\begin{figure}
\begin{centering}
\includegraphics[width=1\columnwidth]{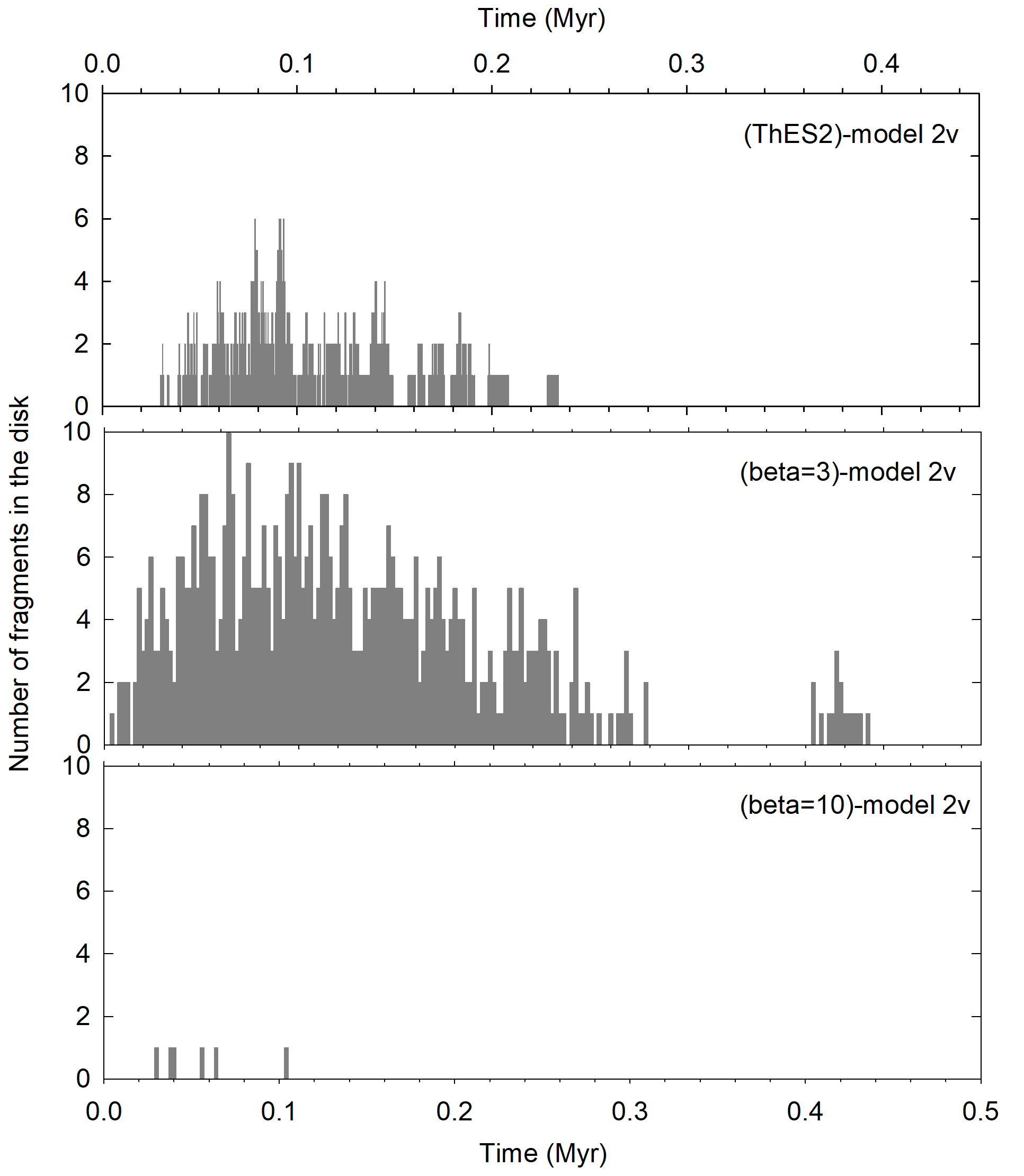}
\par\end{centering}
\caption{\label{fig:11} Comparison of the number of fragments at a given time instance in the disks of (ThES2)-model~2v (top panel), (beta=3)-model~2v (middle panel), and  (beta=10)-model~2v (bottom panel). The time is counted from the instance of disk formation.}
\end{figure}

Figure~\ref{fig:11} presents a comparison of the number of fragments in the disk of (ThES2)-model~2v (top panel), (beta=3)-model~2v (middle panel), and (beta=10)-model~2v (bottom panel). The disk in the $\beta=30$ model did not fragment.  Clearly, the $\beta=3$ model produces too many fragments and the $\beta=10$ model too few fragments as compared to the ThES2 model. The number of fragments reduces in the $\beta$-models with increasing $\beta$-value, as was also found in other studies of disk fragmentation using the $\beta$-cooling scheme \citep[e.g.,][]{Meru2011}. 
The general trend of decreasing strength of gravitational instability with increasing $\beta$-value can be expected -- slower cooling leads to warmer disks and reduced gravitational instability.
Overall, neither of the considered simplified $\beta$-cooling models can reproduce the strength of gravitational instability and fragmentation found in models with a more sophisticated thermal evolution scheme. 

\begin{figure}
\begin{centering}
\includegraphics[width=1\columnwidth]{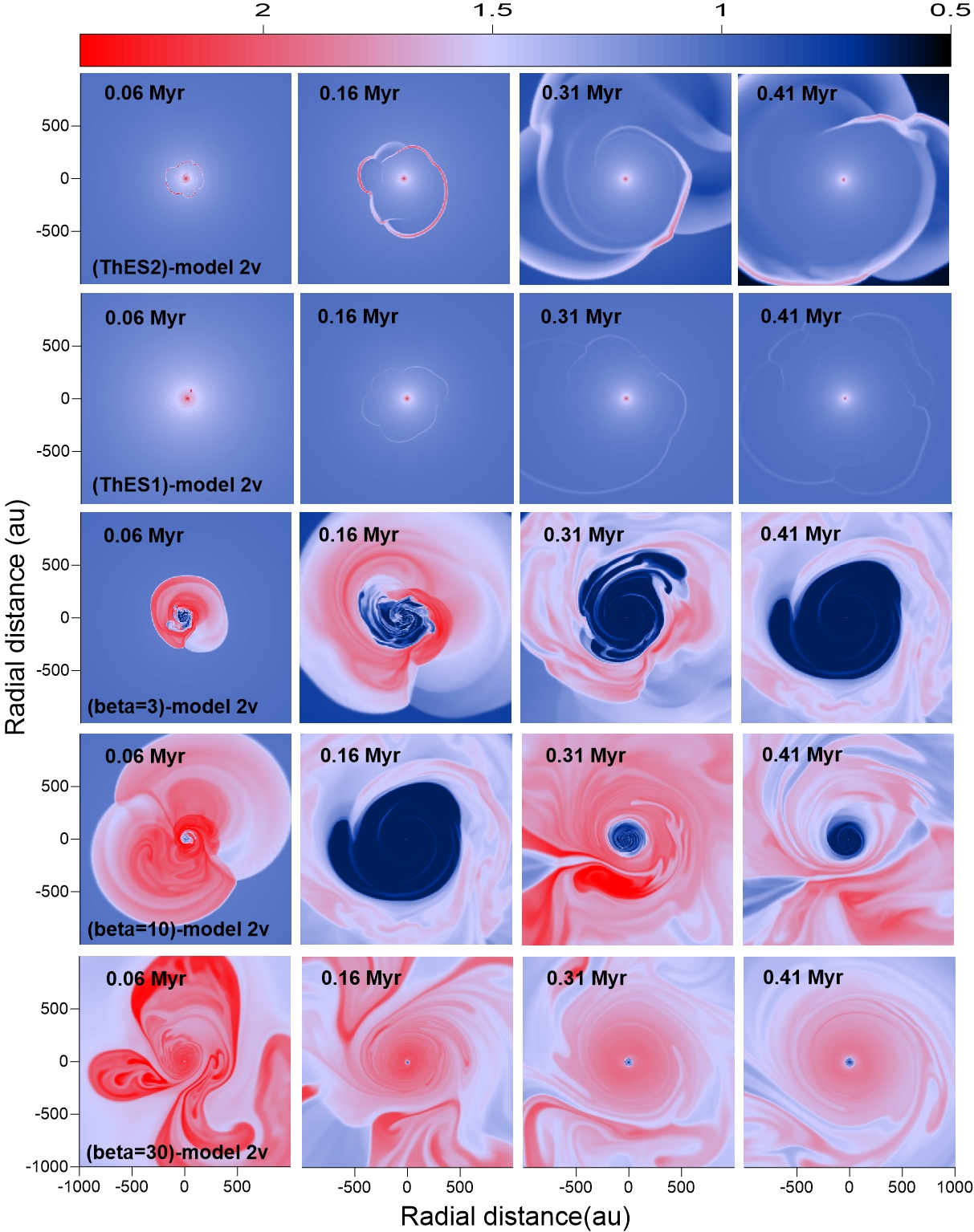}
\par\end{centering}
\caption{\label{fig:12} Comparison of gas temperature distributions in models with different thermal evolution schemes. In particular, the first and second rows present the gas temperatures in (ThES2)-model 2v and (ThES1)-model~2v, respectively. The third, fourth, and last rows show the gas temperatures in (beta=3)-model~2v, (beta=10)-model~2v, and (beta=30)-model~2v, respectively. Note that in the $\beta$-models the gas and dust temperatures are the same.  The time is counted from the instance of disk formation. The scale bar is in log Kelvin.}
\end{figure}

\begin{figure}
\begin{centering}
\includegraphics[width=1\columnwidth]{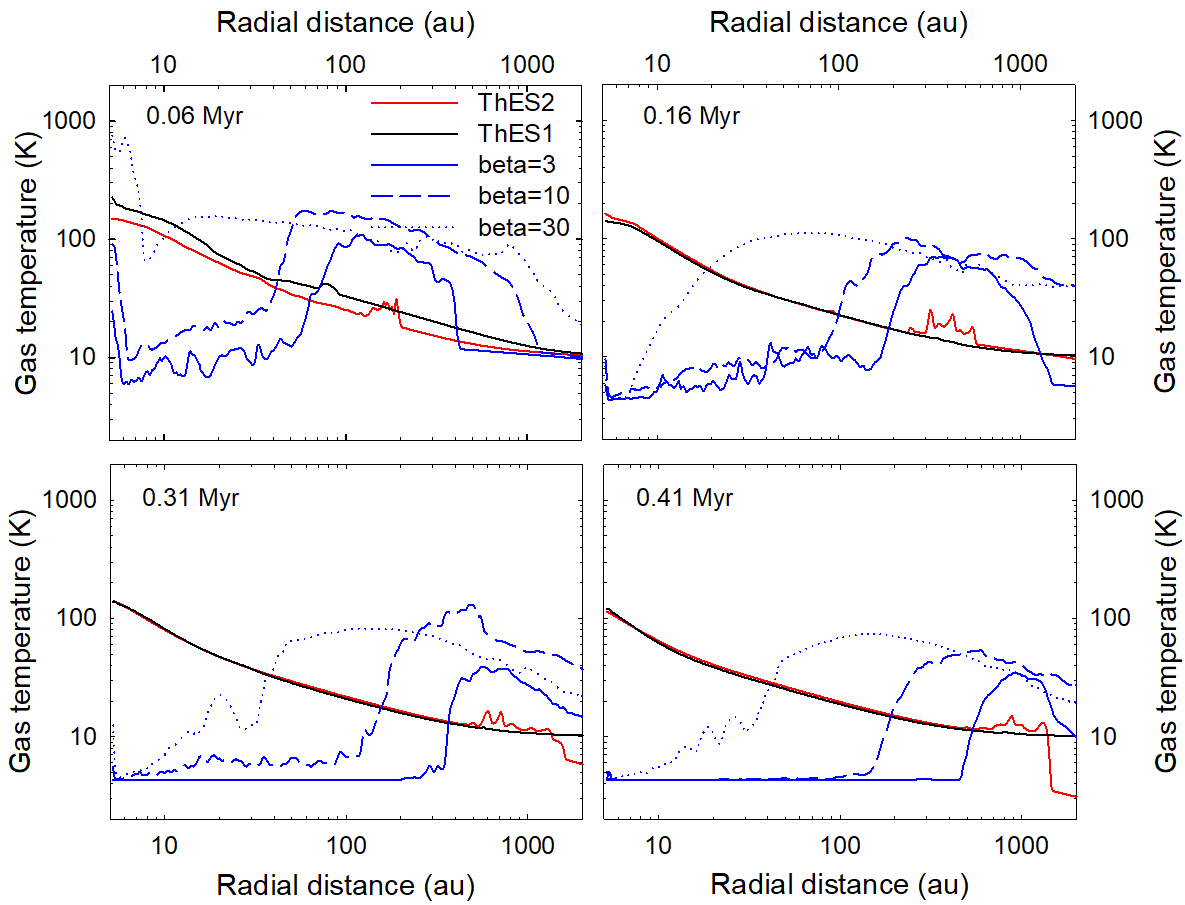}
\par\end{centering}
\caption{\label{fig:14} Comparison of the azimuthally averaged gas temperatures in models with different thermal evolution schemes. In particular, the red and black solid lines present the gas temperature profiles for ThES2 and ThES1 in model~2v, respectively. The blue solid, dashed, and dotted lines show the profiles for model~2v with $\beta=3$, $\beta=10$, and $\beta=30$, respectively. }
\end{figure}

Finally, in the first and second rows of Figure~\ref{fig:12} we present the spatial distribution of gas temperatures in (ThES2)-model~2v and (ThES1)-model~2v, respectively. The other three rows show the gas temperatures in the $\beta$-cooling models. Each column corresponds to a specific age of the disk. The gas temperatures in the $\beta$-models are strikingly different from those of gas and dust in the ThES2 model. In fact, the inner disk regions in the former models are often colder than the periphery, which is a direct consequence of decreasing cooling time with decreasing radial distance for a spatially constant $\beta$-parameter.   This trend is corroborated in  Figure~\ref{fig:14} showing the azimuthally averaged gas temperature profiles for the same models and at the same evolutionary times as in Figure~\ref{fig:12}. The mismatch between the $\beta$-models and more sophisticated thermal evolution schemes is significant.
Note that we had to impose an absolute lower limit on the gas temperature (4~K) in the $\beta$-models to avoid overcooling in the inner disk regions.

\begin{figure}
\begin{centering}
\includegraphics[width=1\columnwidth]{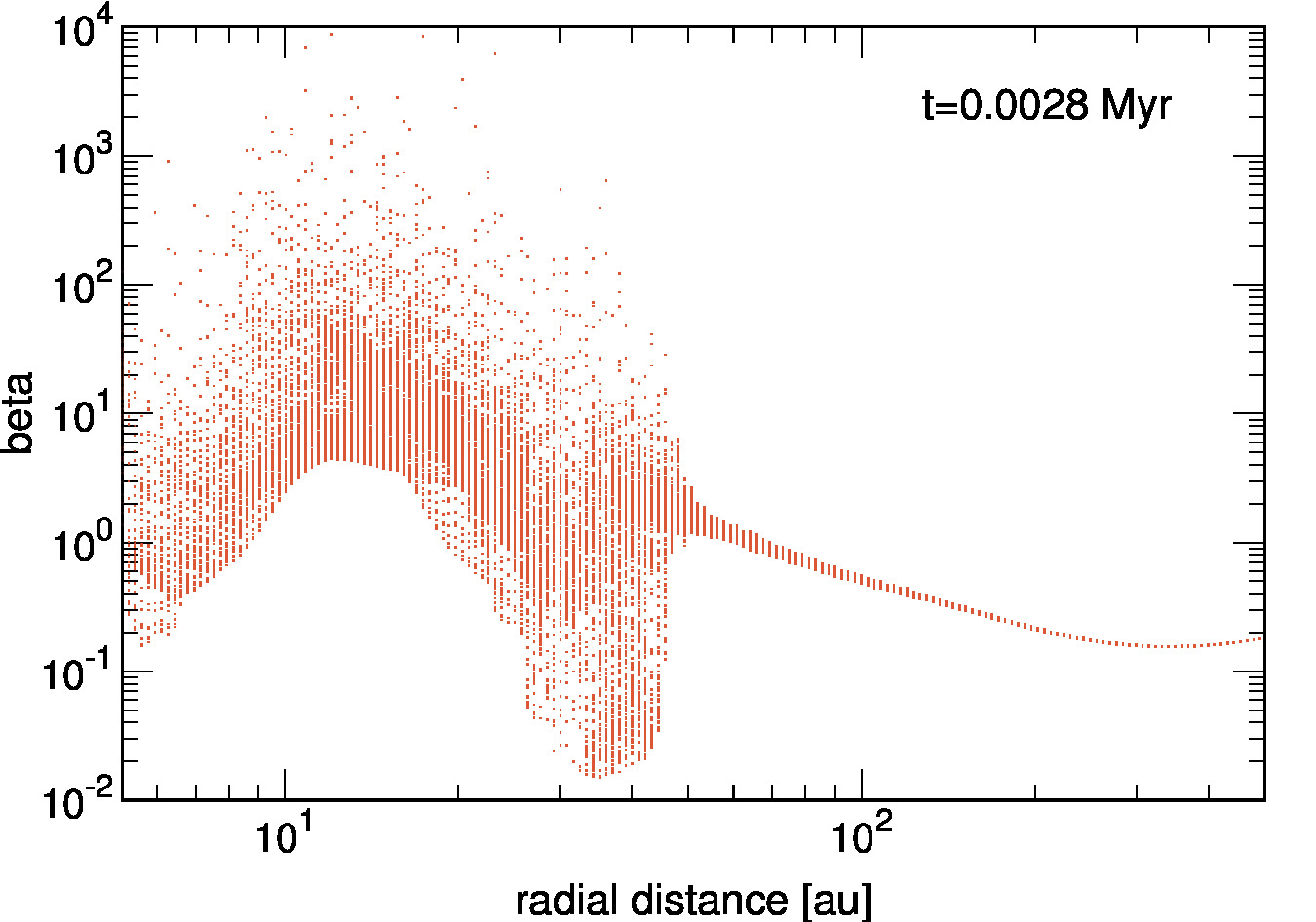}
\par\end{centering}
\caption{\label{fig:13} Values of the $\beta$-parameter calculated in (ThES1)-model~2 at 2.8~kyr after the disk formation instance. Each red dot corresponds to a $\beta$-value in an individual grid cell. Significant azimuthal and radial variations are evident.}
\end{figure}

It may appear that choosing the right $\beta$-value one can achieve a better agreement with the ThES2 model.  This is unlikely because the characteristic cooling time $t_{\rm c}$ can be highly variable both in time and space, meaning that $\beta$ is also variable. We estimated the cooling time in (ThES1)-model~2 as $t_{\rm c}=e/\Lambda$ and confirmed that $\beta=t_{\rm c} \Omega$ varies by more than an order of magnitude both radially and azimuthally within the disk extent, as Fig.~\ref{fig:13} demonstrates.

\subsection{The case with irradiation}
In this section, we present the results for the $\beta$-models that take stellar and background irradiation into account. As compared to the previous section, we dropped the model with $\beta=30$, because the models with increasingly higher $\beta$-values demonstrated progressively worse agreement with the ThES2 thermal scheme. Instead, we considered smaller values of $\beta$, which showed better agreement.

\begin{figure}
\begin{centering}
\includegraphics[width=1\columnwidth]{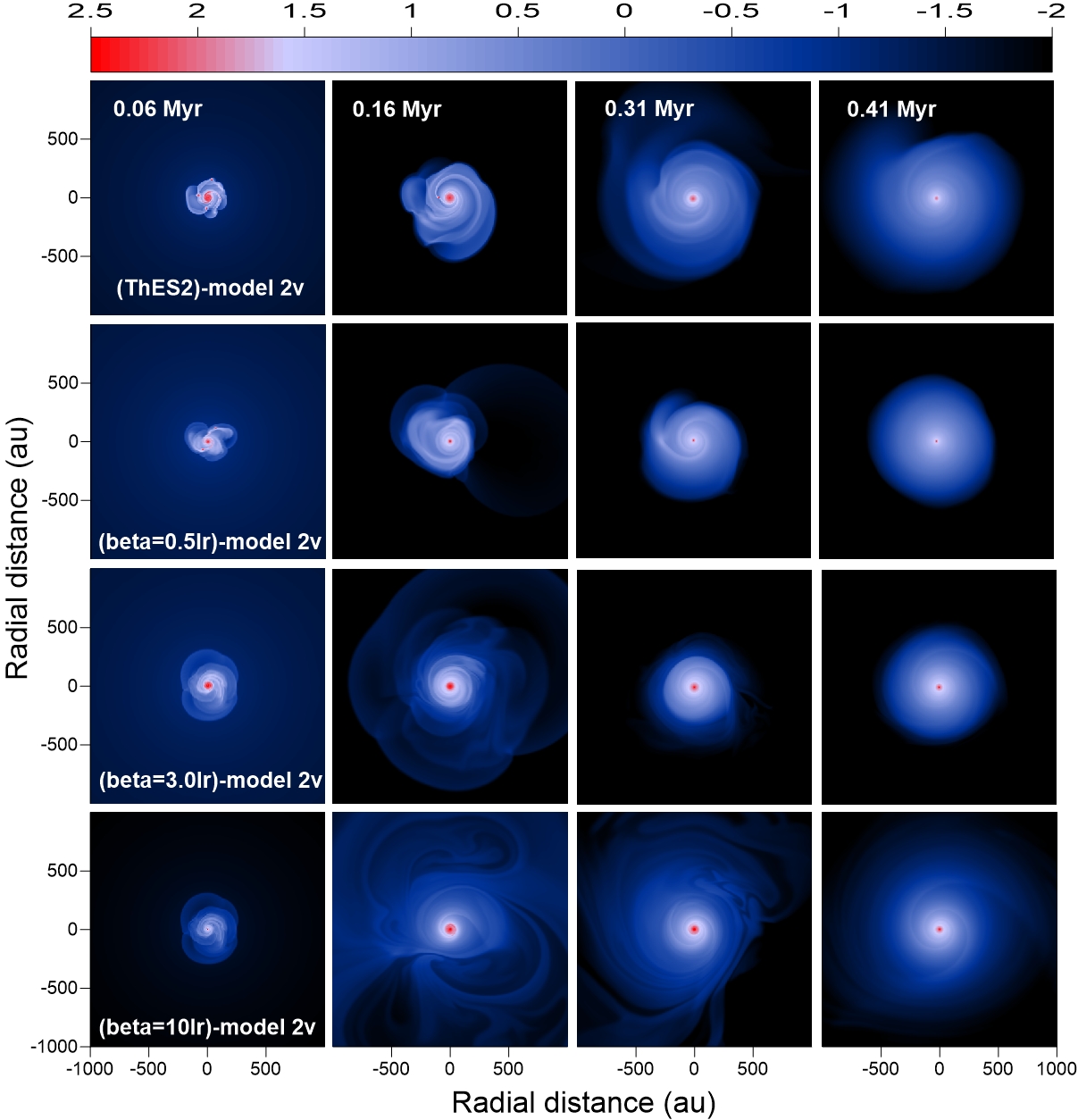}
\par\end{centering}
\caption{\label{fig:15} Comparison of gas surface density distributions in model~v2 with ThES2 and in $\beta$-models that take stellar and background irradiation into account. In particular, the first row corresponds to (ThES2)-model 2v, second row -- to (beta=0.5Ir)-model~2v, third row -- to (beta=3Ir)-model~2v, and bottom row -- to (beta=10Ir)-model~2v. The red circles outline the radial distance beyond which the azimuthally averaged angular velocity deviates from the Keplerian rotation by more than 10\%.
The time is counted from the instance of disk formation. The scale bar is in log g~cm$^{-2}$. }
\end{figure}

Figure~\ref{fig:15} shows the gas surface density distributions in the inner $2000\times2000$~au$^2$ box for (ThES2)-model~2v (top row) and three $\beta$-models with the $\beta$-values set equal to 0.5,  3, and 10. The $\beta$-models have the same initial parameters of prestellar cores and the same value of the viscous $\alpha$ parameter ($10^{-2}$) as in (ThES2)-model~2v. A comparison of Figures~\ref{fig:8} and \ref{fig:15} shows that the $\beta$-models with irradiation can better reproduce the disk evolution obtained in the ThES2 thermal scheme. In particular, models with $\beta=3$ and especially with $\beta=0.5$ possess the disk structure that is similar to what was obtained in (ThES2)-model~2v. As the red circles demonstrate, the near-Keplerian rotation is established in the $\beta=0.5$ model throughout most of the disk extent and at all considered evolutionary times, much like in (ThES2)-model~2v. Nevertheless, some differences can still be noticed, for example, a more compact disk in the $\beta$-models at later evolutionary stages and a more diffused disk in the $\beta=3.0$ model in the early evolution. The $\beta=10$  model (and higher-$\beta$ models) cannot reproduce the disk structure obtained with the most sophisticated thermal evolution scheme, irrespective of whether or not we take irradiation into account.

\begin{figure}
\begin{centering}
\includegraphics[width=1\columnwidth]{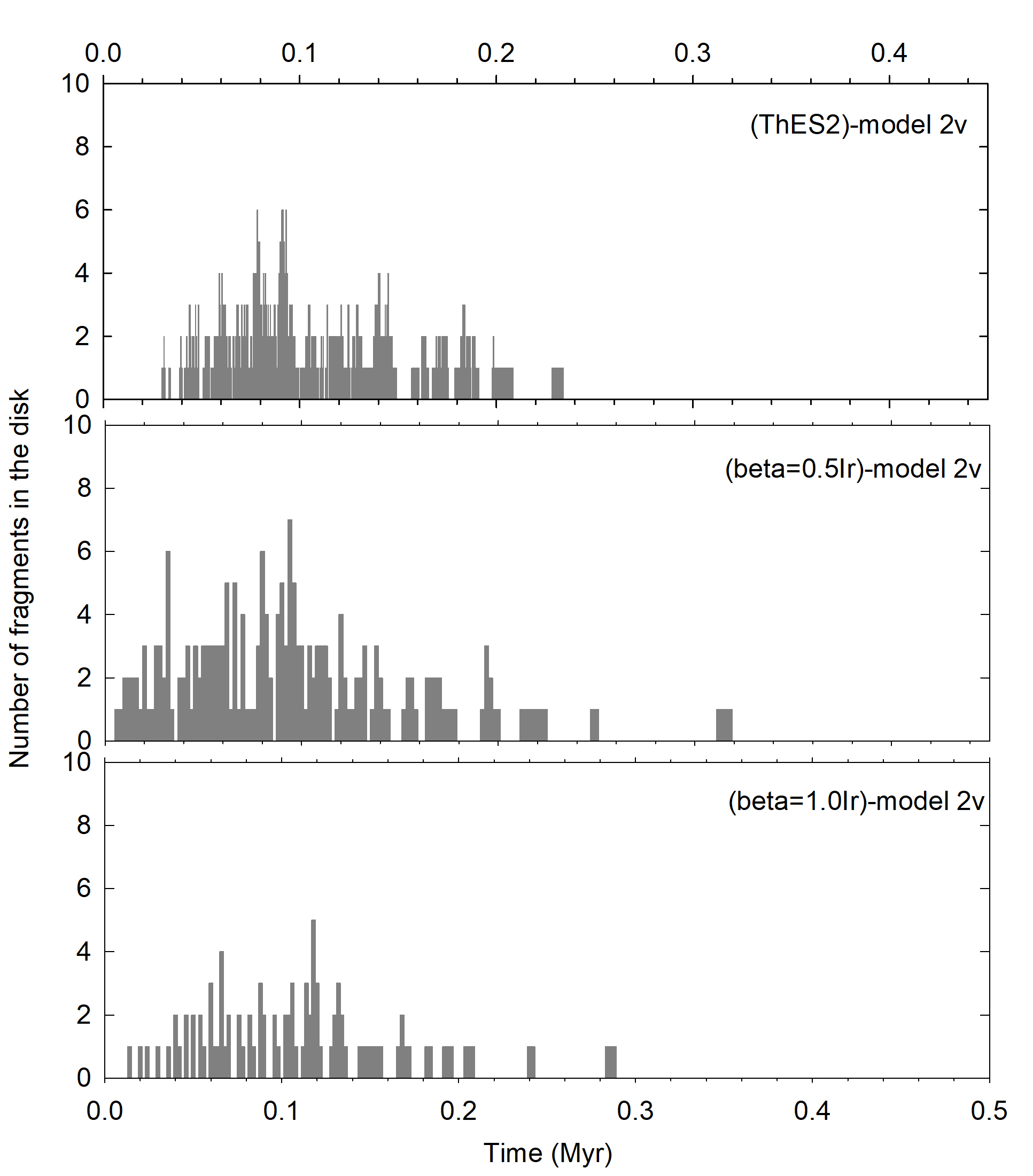}
\par\end{centering}
\caption{\label{fig:16}  Comparison of the number of fragments at a given time instance in the disks of (ThES2)-model~2v (top panel), (beta=0.5Ir)-model~2v (middle panel), and  (beta=1.0Ir)-model~2v (bottom panel). The time is counted from the instance of disk formation.}
\end{figure}

 To corroborate our conclusions, we show in Figure~\ref{fig:16} the number of fragments formed via gravitational fragmentation in the disk of (ThES2)-model~2v and in the disk of $\beta=0.5$ and $\beta=1.0$ models with stellar and background irradiation. We note that models with higher values of $\beta$ did not show disk fragmentation, including the $\beta=3.0$ model considered previously. What concerns the number of fragments, the $\beta=0.5$~model can best reproduce disk fragmentation in (ThES2)-model~2v. The $\beta=1.0$ model forms too few fragments compared to (ThES2)-model~2v. Nevertheless, in both $\beta$-models disk fragmentation starts earlier and ends later than in the model with the most sophisticated thermal evolution scheme.

\begin{figure}
\begin{centering}
\includegraphics[width=1\columnwidth]{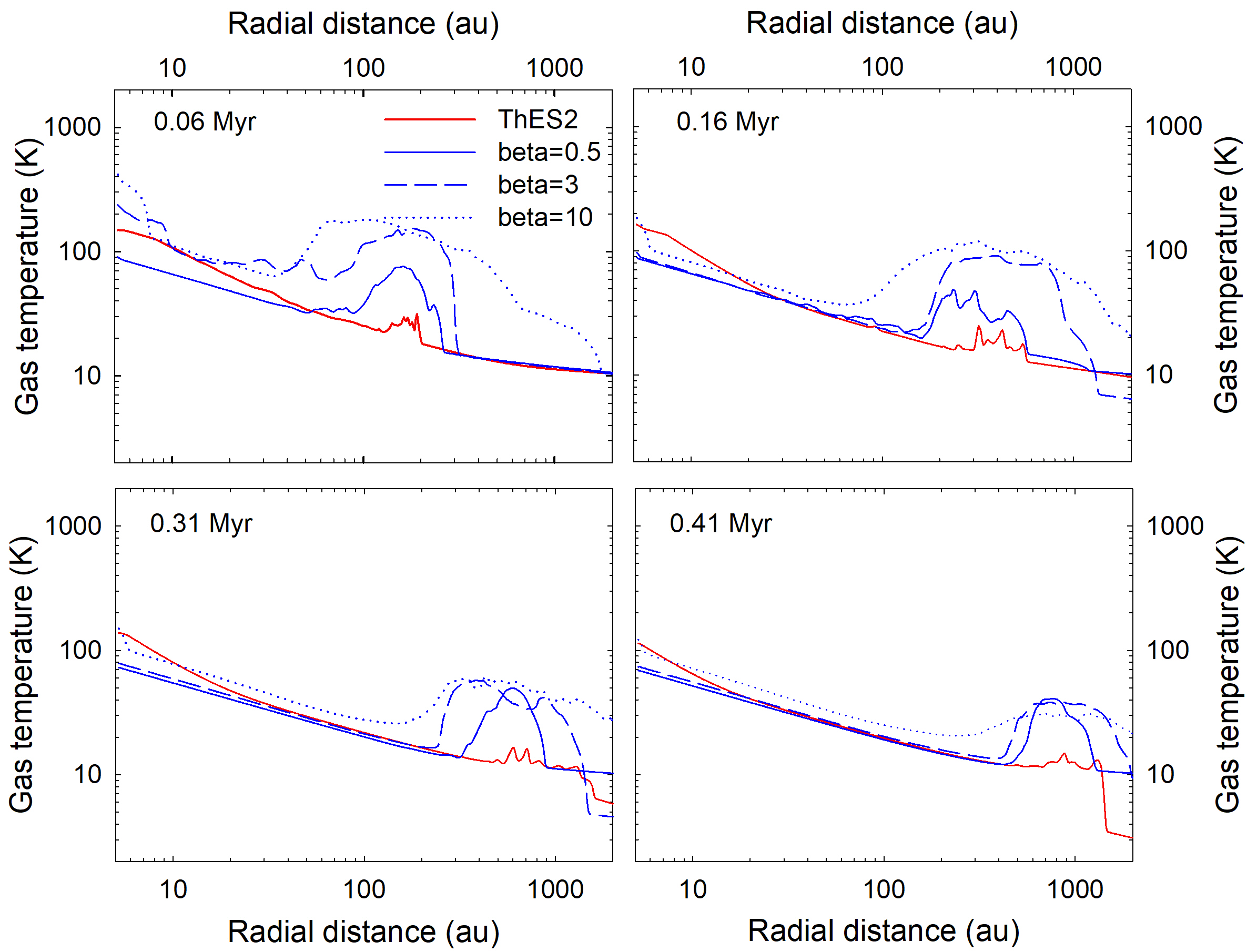}
\par\end{centering}
\caption{\label{fig:17} Comparison of the azimuthally averaged gas temperatures in (TheS2)-model~2v (red lines) and in $\beta$-models with irradiation. In particular, the blue solid, dashed, and dotted lines show the profiles for model~2v with $\beta=0.5$, $\beta=3$, and $\beta=10$, respectively.}
\end{figure}

Finally, Figure~\ref{fig:17} presents the azimuthally
averaged radial profiles of gas temperature in (ThES2)-model~2v and in three $\beta$-models with irradiation. Clearly, the $\beta=0.5$~model can best reproduce the radial temperature profile in the most sophisticated thermal scheme. Nevertheless, the inner disk regions are colder in this $\beta$-model as compared to (TheS2)-model~2v. Strong $\beta$-cooling in the inner fast-rotating disk regions completely overwhelms disk viscous heating in the lowest-$\beta$ model. 
Interestingly, all $\beta$-models show a local increase in the gas temperature in the outer disk regions and in the inner envelope, a feature that is also present in the ThES2 scheme, but is absent in the ThES1 scheme (see Fig.~\ref{fig:14}). This may be related to strong compressional heating due to PdV work of infalling envelope material and reduced $\beta$-cooling in the outer regions with slow rotation. The amplitude of the temperature jump in the $\beta$-models is a factor of several higher than in (ThES2)-model~2v. 

We conclude that $\beta$-models with irradiation can much better reproduce the thermal state obtained in the ThES2 scheme than the $\beta$-models without irradiation, but the agreement is still not acceptable. We note that the required computational resources for the ThES2 scheme are significantly higher those for the ThES1 and $\beta$-schemes. Therefore, the former scheme is advisable for simulations where gas and dust temperature decoupling is expected to be of particular importance. In other situations, the ThES1 scheme may be a method of choice due to its relatively easy coding and moderate computational resources required.

\section{Summary and Discussion}
\label{Conclusions}

Here we explored numerically the global long-term evolution of protostellar disks with different cooling-heating schemes. For this purpose, we used three approaches to describing the thermal balance in the disk: a simplistic $\beta$-cooling scheme with and without stellar and background irradiation considered, a more realistic scheme with similar gas and dust temperatures (ThES1), and a sophisticated scheme with separate gas and dust temperatures (ThES2). The latter scheme can also be applied to low-metallicity protostellar disks.

The adopted thermal schemes were tested on global disk models computed in the thin-disk limit without and with turbulent viscosity using the Shakura \& Sunyaev $\alpha$-parameterization.
The main results can be summarized as follows.

-- In the ThES2 scheme, the gas and dust temperatures are similar in the inner high-density regions of the disk, but may significantly deviate from each other in the vicinity of the disk outer edge. In this low-density region a pronounced decoupling between the dust and gas temperatures develops thanks to additional compressional heating of the gas caused by the infalling envelope material and because of slow collisional energy exchange between gas and dust in the low-density environment. The gas temperature may exceed that of dust by tens or even hundreds of Kelvin. 

-- In the outer circumdisk environment occupied by a rarefied envelope the gas temperature also decouples from that of dust, but in the opposite direction -- the gas temperature drops below that of dust. This effect was also reported in \citet{Pavlyuchenkov2015} and \citet{Bate2018}.

-- The global disk evolution is weakly sensitive to decoupling of gas and dust temperatures. Gravitational instability in the case of separate gas and dust temperatures (ThES2) is only slightly stronger than in the case of similar gas and dust temperatures (ThES1), as indicated by higher Fourier amplitudes, and gravitational fragmentation is slightly more frequent. Overall, separate gas and dust temperatures do not cause qualitative changes to the disk evolution, which appears to be more sensitive to the presence or absence of turbulent viscosity in the disk.

-- Decoupling of gas and dust temperatures may nevertheless be of significance for the chemical evolution and dust growth. An increase in the gas temperature to more than a hundred Kelvin could launch gas phase reactions in the disk outer regions that otherwise may be dormant. Moreover, decoupling of gas and dust temperatures may also facilitate the growth of icy mantles on cold dust particles, if volatile species become oversaturated in the warm gas environment in the vicinity of the disk outer edge. The decoupled gas and dust temperatures may also affect the disk mass estimates.
 
--  Simplistic constant-$\beta$ models without irradiation fail to reproduce the disk evolution in more sophisticated thermal models with or without separate gas and dust temperatures. We attribute this to the intrinsic variability of the $\beta$-parameter both in time and space. $\beta$-cooling models with stellar and background irradiation taken into account can better match the dynamical and thermal evolution obtained in the ThES2 thermal scheme, particularly for $\beta\approx 0.5-1.0$, but the agreement is still incomplete.

In the future, it will be interesting to consider cases when the gas and dust temperatures decouple through the bulk of the disk, and not only in the disk outermost regions and in the envelope. This may occur either at metallicities lower than Solar or when the dust-to-gas ratio drops below the canonical 1:100 value in otherwise solar-metallicity disks.


\section*{Acknowledgements}

We are thankful to the anonymous referee for constructive comments that helped to improve the manuscript.
E.I.V. and M.G. acknowledge support from the Austrian Science Fund (FWF) under research grant 
P31635-N27. K.O and R.M acknowledge support work by MEXT/JSPS KAKENHI Grant Number17H01102, 17H02869, 17H06360.
The simulations were performed on the Vienna Scientific Cluster.

\appendix

\section{Line-averaged escape probability of HD}
\label{App:HD-escape}
The HD line cooling rate is reduced by the effect of self-absorption when the column density of HD in the disk is large. 
The photon escape probability for self-absorption in individual transitions is gevin by 
\begin{equation}
\beta_{\rm HD,ul} = \frac{1-\mathrm{e}^{-\tau_{\rm ul}}}{\tau_{\rm ul}}~,
\end{equation}
where $\tau_{\rm ul}$ is the optical depth for line emission given by the same formula of Equation~(\ref{Eq:line-tau}). 
The photon escape probability is a function of the column density of HD and the gas temperature because 
the optical depth for line emission depends on them as you can see from Equations~(\ref{Eq:line-tau}) and~(\ref{Eq:thermal-velocity}). 
We treat four levels in the range of rotational levels $0<{\rm J}<3$. 
The level energies and the spontaneous radiative decay rates are referred from \cite{Dalgarno1972} and \cite{Abgrall1982}, respectively. 
We only consider helium impact as the collisional de-excitation process and its rate coefficients are taken from \cite{Galli1998}. 
The cooling rate of HD line emission is calculated by counting level populations from the statistical balance among four levels. 
We calculate the cooling rate in the range of the column density of HD $10^{15}~{\rm cm^{-2}}<N_{\rm HD}<10^{25}~{\rm cm^{-2}}$ and 
the gas temperature $30~{\rm K}<T_{\rm g}<3000~{\rm K}$. 
We fit the line-averaged escape probability getting from the results in a functional form as 
\begin{equation}
    \overline{\beta}_{\rm esc,HD} = \frac{1}{\left(1+N_{\rm HD}/N_{\rm c}\right)^{\alpha}}~,
    \label{Eq:HD_escape}
\end{equation}
where
\begin{equation}
    \alpha = a_{0} + a_{1}T_{\rm g} + a_{2}T_{\rm g}^{2} + a_{3}T_{\rm g}^{3} +
    a_{4}T_{\rm g}^{4} + a_{5}T_{\rm g}^{5}~,
    \label{Eq:A2}
\end{equation}
and
\begin{eqnarray}
    \mathrm{log}~N_{\rm c} = &b_{0}& + b_{1}\mathrm{log}~T_{\rm g} + b_{2}\left(\mathrm{log}~T_{\rm g}\right)^{2} + 
    b_{3}\left(\mathrm{log}~T_{\rm g}\right)^{3} \notag \\
    &+& b_{4}\left(\mathrm{log}~T_{\rm g}\right)^{4} + b_{5}\left(\mathrm{log}~T_{\rm g}\right)^{5}~.
    \label{Eq:A3}
\end{eqnarray}
The fitting coefficients in Equations~(\ref{Eq:A2}) and~(\ref{Eq:A3}) are given 
in Tables~\ref{Table:alpha} and~\ref{Table:Nc}.

\begin{table*}
 \begin{center}
 \caption{The fitting coefficients for $\alpha$}
 \label{Table:alpha}
  \scalebox{0.8}[1.1]{ 
  \begin{tabular}{l c c c c c c} \hline \hline
    temperature (K) & $a_{0}$ & $a_{1}$ & $a_{2}$ & $a_{3}$ & $a_{4}$ & $a_{5}$ \\ \hline
    $T_{\rm g}$ $<$ 1000 & 1.0234 & -9.0445$\times 10^{-4}$ & 3.2614$\times 10^{-6}$ & -5.7367$\times 10^{-9}$
    & 4.9825$\times 10^{-12}$ & -1.6808$\times 10^{-15}$ \\ 
    $T_{\rm g}$ $>$ 1000 & 8.9433$\times 10^{-1}$ & 6.2158$\times 10^{-5}$ & -1.4744$\times 10^{-9}$ 
    & -9.0761$\times 10^{-12}$ & 3.3103$\times 10^{-15}$ & -3.7980$\times 10^{-19}$ \\ \hline
  \end{tabular}
  }
 \end{center}
\end{table*}
\begin{table*}
 \begin{center}
 \caption{The fitting coefficients for $N_{\rm c}$}
 \label{Table:Nc}
  \scalebox{0.9}[1.0]{ 
  \begin{tabular}{c c c c c c} \hline \hline
    $b_{0}$ & $b_{1}$ & $b_{2}$ & $b_{3}$ & $b_{4}$ & $b_{5}$ \\ \hline
    18.892 & -1.4802 & 8.5905$\times 10^{-1}$ & 4.3840$\times 10^{-1}$ 
    & -3.0569$\times 10^{-1}$ & 4.6412$\times 10^{-2}$ \\ \hline
  \end{tabular}
  }
 \end{center}
\end{table*}

\section{Chemical reactions}

In our new cooling-heating scheme described in Section \ref{ThES2}, 
we follow the non-equilibrium chemical evolution. 
Our chemical network is composed of 21 hydrogen reactions and 6 deuterium reactions. 
We describe treated chemical reactions and their rate coefficients in Table \ref{Table:reactions}. 
In reactions from 1 to 20, only the rate coefficients of forward reaction are shown. 
The calculation method of the rate coefficients of reverse reactions is referred from \cite{Matsukoba2019}. 

\begin{table*}
 \begin{center}
 \caption{Chemical reactions}
 \label{Table:reactions}
  \scalebox{0.85}[1.0]{ 
  \begin{tabular}{c l l c} \hline \hline
    Number & Reaction & Rate coefficient ($\mathrm{cm}^{3}~\mathrm{s^{-1}}$) & Reference \\ \hline
    1, 2 & $\mathrm{H} + \mathrm{e} \rightleftharpoons \mathrm{H}^{+} + 2 \mathrm{e}$ & $k_{1} = \mathrm{exp}[-3.271396786 \times 10^{1}$ & {\small \cite{Janev:1987}} \\ 
    & & $~~~~~~ + 1.35365560 \times 10^{1}~\mathrm{ln}~T_{\mathrm{e}} - 5.73932875 \times 10^{0}~(\mathrm{ln}~T_{\mathrm{e}})^{2}$ & \\
    & & $~~~~~~ + 1.56315498 \times 10^{0}~(\mathrm{ln}~T_{\mathrm{e}})^{3} - 2.87705600 \times 10^{-1}~(\mathrm{ln}~T_{\mathrm{e}})^{4}$ & \\
    & & $~~~~~~ + 3.48255977 \times 10^{-2}~(\mathrm{ln}~T_{\mathrm{e}})^{5} - 2.63197617 \times 10^{-3}~(\mathrm{ln}~T_{\mathrm{e}})^{6}$ & \\
    & & $~~~~~~ + 1.11954395 \times 10^{-4}~(\mathrm{ln}~T_{\mathrm{e}})^{7} - 2.03914985 \times 10^{-6}~(\mathrm{ln}~T_{\mathrm{e}})^{8}]$ & \\
    3, 4 & $\mathrm{H}^{-} + \mathrm{H} \rightleftharpoons \mathrm{H}_{2} + \mathrm{e}$ & 
       $k_{3} = 1.3500\times10^{-9}( T_{\mathrm{g}}^{9.8493\times10^{-2}}+3.2852\times10^{-1}T_{\mathrm{g}}^{5.5610\times10^{-1}}$ & {\small \cite{Kreckel:2010}} \\    
    & & $~~~~~~ +2.7710\times10^{-7}T_{\mathrm{g}}^{2.1826})/(1.0+6.1910\times10^{-3}T_{\mathrm{g}}^{1.0461}$ & \\
    & & $~~~~~~ +8.9712\times10^{-11}T_{\mathrm{g}}^{3.0424}+3.2576\times10^{-14}T_{\mathrm{g}}^{3.7741})$ & \\    
    5, 6 & $\mathrm{H}_{2} + \mathrm{e} \rightleftharpoons 2 \mathrm{H} + \mathrm{e}$ & $k_{5} = k_{5, \mathrm{H}}^{1-a} k_{5, \mathrm{L}}^{a} $ & \\
    & & $k_{5, \mathrm{H}} = 1.91 \times 10^{-9} T_{\mathrm{g}}^{0.136} \mathrm{exp}\Bigl( - 53407.1/T_{\mathrm{g}} \Bigr)$ & {\small \cite{Trevisan:2002}} \\
    & & $k_{5, \mathrm{L}} = 4.49 \times 10^{-9} T_{\mathrm{g}}^{0.11} \mathrm{exp}\Bigl( - 101858/T_{\mathrm{g}} \Bigr)$ & \\
    & & $a = \left( 1 + n_{\mathrm{H}} / n_{\mathrm{crit}} \right)^{-1}$ & \\        
    & & $n_{\mathrm{crit}} = \left[y(\mathrm{H})/n_{\mathrm{crit}}(\mathrm{H}) + 2y(\mathrm{H}_{2})/n_{\mathrm{crit}}(\mathrm{H}_{2}) 
       + y(\mathrm{He})/n_{\mathrm{crit}}(\mathrm{He})\right]^{-1}$ & \\            
    & & $\mathrm{log}~(n_{\mathrm{crit}}(\mathrm{H})) = 3 - 0.416~\mathrm{log}~(T_{\mathrm{g}}/10^{4}) - 0.372 \left[ \mathrm{log}~(T_{\mathrm{g}}/10^{4}) \right]^{2}$ & \\                
    & & $\mathrm{log}~(n_{\mathrm{crit}}(\mathrm{H}_{2})) = 4.845 - 1.3~\mathrm{log}~(T_{\mathrm{g}}/10^{4}) + 1.62 \left[ \mathrm{log}~(T_{\mathrm{g}}/10^{4}) \right]^{2}$ & \\                
    & & $\mathrm{log}~(n_{\mathrm{crit}}(\mathrm{He})) = 5.0792 \left[ 1 - 1.23 \times 10^{-5} (T_{\mathrm{g}} - 2000) \right]$ & \\            
    7, 8 & $3 \mathrm{H} \rightleftharpoons \mathrm{H}_{2} + \mathrm{H}$ & $k_{7} = 7.7\times10^{-31}T_{\mathrm{g}}^{-0.464}$ & {\small \cite{Glover:2008-3}} \\
    9, 10 & $2 \mathrm{H} + \mathrm{H}_{2} \rightleftharpoons 2 \mathrm{H}_{2}$ & $k_{9} = k_{7}/8$ & {\small \cite{Palla:1983}} \\
    11, 12 & $\mathrm{H}^{-} + \mathrm{H}^{+} \rightleftharpoons 2 \mathrm{H}$ & $k_{11} = 2.4 \times 10^{-6} T_{\mathrm{g}}^{-0.5} \Bigl( 1.0 + T_{\mathrm{g}}/20000 \Bigr)$ & {\small \cite{Croft:1999}} \\
    13, 14 & $\mathrm{H}^{+} + \mathrm{e} \rightleftharpoons \mathrm{H} + \gamma$ & 
       $k_{13} = 2.753 \times 10^{-14} \Bigl( 315614/T_{\mathrm{g}} \Bigr)^{1.5} \Bigl[ 1.0 + \Bigl( 115188/T_{\mathrm{g}} \Bigr)^{0.407} \Bigr]^{-2.242}$ & {\small \cite{Ferland:1992}} \\
    15, 16 & $\mathrm{H} + \mathrm{e} \rightleftharpoons \mathrm{H}^{-} + \gamma$ & 
       $k_{15} = \mathrm{dex}[-17.845 + 0.762 \mathrm{log}~T_{\mathrm{g}} + 0.1523 (\mathrm{log}~T_{\mathrm{g}})^{2}$ & {\small \cite{Wishart:1979}} \\
    & & $~~~~~~ - 0.03274 (\mathrm{log}~T_{\mathrm{g}})^{3}]$ ~~~~~~~~~ ($T_{\mathrm{g}} < 6000~\mathrm{K}$) & \\
    & & $~~~~ = \mathrm{dex}[-16.4199 + 0.1998 (\mathrm{log}~T_{\mathrm{g}})^{2} - 5.447 \times 10^{-3} (\mathrm{log}~T_{\mathrm{g}})^{4}$ & \\
    & & $~~~~~~ + 4.0415 \times 10^{-5} (\mathrm{log}~T_{\mathrm{g}})^{6}]$ ~ ($T_{\mathrm{g}} > 6000~\mathrm{K}$) & \\
    17, 18 & $\mathrm{H}_{2} + \mathrm{He} \rightleftharpoons 2 \mathrm{H} + \mathrm{He}$ & $k_{17} = k_{17, \mathrm{H}}^{1-a} k_{17, \mathrm{L}}^{a} $ & \\    
    & & $k_{17, \mathrm{H}} =  \mathrm{dex}[-1.75~\mathrm{log}T_{\mathrm{g}} -2.729 - 23474/T_{\mathrm{g}}]$ & {\small \cite{Dove:1987}} \\
    & & $k_{17, \mathrm{L}} = \mathrm{dex}[3.801~\mathrm{log}T_{\mathrm{g}} -27.029 - 29487/T_{\mathrm{g}}]$ & \\
    19, 20 & $2 \mathrm{H} \rightleftharpoons \mathrm{H}^{+} + \mathrm{e} + \mathrm{H}$ & 
        $k_{19} = 1.2¥\times10^{-17}T_{\mathrm{g}}^{1.2}~\mathrm{exp}\left( -\frac{157800}{T_{\mathrm{g}}} \right)$ & {\small \cite{Lenzuni:1991}} \\         
    21 & $2\mathrm{H} + \mathrm{grain} \rightarrow \mathrm{H}_{2}$ & $k_{21} = 6.0\times10^{-17}\left(T_{\mathrm{g}}/300~\mathrm{K}\right)^{1/2}f_{a}$ & {\small \cite{Tielens1985}} \\
    & & $~~~~\times\left[1+4.0\times10^{-2}\left(T_{\mathrm{g}}+T_{\mathrm{d}}\right)^{1/2}+2.0\times10^{-3}T_{\mathrm{g}}+8.0\times10^{-6}T_{\mathrm{g}}^{2}\right]^{-1}\times Z/Z_{\mathrm{local}}$ & \\
    & & $f_{a}=\left\{1+\mathrm{exp}\left[7.5\times10^{2}\left(1/75-1/T_{\mathrm{d}}\right)\right]\right\}^{-1}$ & \\
    22 & $\mathrm{D} + \mathrm{H}^{+} \rightarrow \mathrm{D}^{+} + \mathrm{H}$ & 
       $k_{22} = 2.0\times10^{-10}T_{\mathrm{g}}^{0.402}\mathrm{exp}\left(-37.1/T_{\mathrm{g}}\right)-3.31\times10^{-17}T_{\mathrm{g}}^{1.48}~~~\left(T_{\mathrm{g}}<2\times10^{5}~\mathrm{K}\right)$ & {\small \cite{Savin2002}} \\
    & & $~~~~= 3.44\times10^{-10}T_{\mathrm{g}}^{0.35}~~~~~~~~~~~~~~~~~~~~~~~~~~~~~~~~~~~~~~~~~~~~~~~~~\left(T_{\mathrm{g}}>2\times10^{5}~\mathrm{K}\right)$ & \\
    23 & $\mathrm{D}^{+} + \mathrm{H} \rightarrow \mathrm{D} + \mathrm{H}^{+}$ & 
       $k_{23} = 2.06\times10^{-10}T_{\mathrm{g}}^{0.396}\mathrm{exp}\left(-33/T_{\mathrm{g}}\right)+2.03\times10^{-9}T_{\mathrm{g}}^{-0.332}$ & {\small \cite{Savin2002}} \\
    24 & $\mathrm{D} + \mathrm{H}_{2} \rightarrow \mathrm{HD} + \mathrm{H}$ & 
       $k_{24} = \mathrm{dex}[-56.4737 + 5.88886\mathrm{log}T_{\mathrm{g}} + 7.19692\left(\mathrm{log}T_{\mathrm{g}}\right)^{2} + 2.25069\left(\mathrm{log}T_{\mathrm{g}}\right)^{3} $ & {\small \cite{Glover:2008-8}} \\
    & & $~~~~~- 2.16903\left(\mathrm{log}T_{\mathrm{g}}\right)^{4} + 0.317887\left(\mathrm{log}T_{\mathrm{g}}\right)^{5}]~~~\left(T_{\mathrm{g}}<2000~\mathrm{K}\right)$ & \\
    & & $~~~~= 3.17\times10^{-10}\mathrm{exp}\left(-5207/T_{\mathrm{g}}\right)~~~~~~~~~~~~~~\left(T_{\mathrm{g}}>2000~\mathrm{K}\right)$ & \\
    25 & $\mathrm{D}^{+} + \mathrm{H}_{2} \rightarrow \mathrm{HD} + \mathrm{H}^{+}$ & 
       $k_{25} = \left[0.417+0.846\mathrm{log}T_{\mathrm{g}}-0.137\left(\mathrm{log}T_{\mathrm{g}}\right)^{2}\right]\times10^{-9}$ & {\small \cite{Gerlich1982}} \\
    26 & $\mathrm{HD} + \mathrm{H} \rightarrow \mathrm{H}_{2} + \mathrm{D}$ & 
       $k_{26} = 5.25\times10^{-11}\mathrm{exp}\left(-4430/T_{\mathrm{g}}\right)~~~~~~~~~~~~~~~~~~~~~\left(T_{\mathrm{g}}<200~\mathrm{K}\right)$ & {\small \cite{Shavitt1959}} \\
    & & $~~~~= 5.25\times10^{-11}\mathrm{exp}\left(-4430/T_{\mathrm{g}} + 173900/T_{\mathrm{g}}^{2}\right)~~~\left(T_{\mathrm{g}}>200~\mathrm{K}\right)$ & \\
    27 & $\mathrm{HD} + \mathrm{H}^{+} \rightarrow \mathrm{D}^{+} + \mathrm{HD}$ & $k_{27} = 1.1\times10^{-9}\mathrm{exp}\left(-488/T_{\mathrm{g}}\right)$ & {\small \cite{Gerlich1982}} \\ \hline
    \multicolumn{4}{l}{{\footnotesize Note.--- The temperature $T_{\mathrm{e}}$ is in eV.}} \\
  \end{tabular}
  }
  \\
 \end{center}
\end{table*}


\begin{thebibliography}{}
\bibitem[\protect\citeauthoryear{Abgrall et al.}{1982}]{Abgrall1982}
Abgrall, H., Roueff, E., \& Viala, Y. 1982, A\&AS, 50, 505

\bibitem[\protect\citeauthoryear{Andrews et al.}{2009}]{Andrews2009}
Andrews, S. M., Wilner, D. J., Hughes, A. M., Qi, C., \&  Dullemond, C. P. 2009, ApJ, 700, 1502

\bibitem[\protect\citeauthoryear{Baehr \& Klahr}{2015}]{Baehr2015}
Baehr, H. \& Klahr, H. 2015, ApJ, 814, 155

\bibitem[\protect\citeauthoryear{Basu}{1997}]{Basu1997}
Basu, S. 1997, ApJ, 485, 240

\bibitem[\protect\citeauthoryear{Bate \& Keto}{2015}]{Bate2015}
Bate, M. R., \& Keto, E. R. 2015, MNRAS, 449, 2643

\bibitem[\protect\citeauthoryear{Bate}{2018}]{Bate2018}
Bate, M. R. 2018, MNRAS, 475, 5618

\bibitem[\protect\citeauthoryear{Binney \& Tremaine}{1987}]{BT87}
Binney, J., \& Tremaine, S. 1987, Galactic Dynamics ( Princeton: Princeton Univ. Press)

\bibitem[\protect\citeauthoryear{Boley et al.}{2010}]{Boley2010}
Boley, A. C., Hayfield, T., Mayer, L., \& Durisen, R. H. 2010, Icarus, 207, 509

\bibitem[\protect\citeauthoryear{Boss}{2002}]{Boss2002}
Boss, A. P. 2002, ApJ, 576, 462

\bibitem[\protect\citeauthoryear{Boss}{2017}]{Boss2017}
Boss, A. P. 2017, ApJ, 836, 53

\bibitem[\protect\citeauthoryear{Colella \& Woodard}{1984}]{Colella1984}
Colella, P., \& Woodward, P. R. 1984, J. Comput. Phys., 54, 174

\bibitem[\protect\citeauthoryear{Cossins et al.}{2009}]{Cossins2009}
Cossins P., Lodato G., Clarke C. J., 2009, MNRAS, 393, 1157

\bibitem[\protect\citeauthoryear{{Croft}, {Dickinson}  \& {Gadea}}{{Croft}
  et~al.}{1999}]{Croft:1999}
{Croft} H.,  {Dickinson} A.~S.,   {Gadea} F.~X.,  1999, MNRAS, 304, 327

\bibitem[\protect\citeauthoryear{Dalgarno \& Wright}{1972}]{Dalgarno1972}
Dalgarno, A., \& Wright, E. L. 1972, ApJ, 174, L49

\bibitem[\protect\citeauthoryear{Deng et al.}{2017}]{Deng2017}
Deng, H., Mayer, L., Meru, F. 2017, ApJ, 847, 43

\bibitem[\protect\citeauthoryear{Dong et al.}{2016}]{Dong2016}
Dong, R., Vorobyov, E., Pavlyuchenkov, Y., Chiang, E., \& Liu, H. B. 2016, ApJ, 823, 141

\bibitem[\protect\citeauthoryear{{Dove}, {Rusk}, {Cribb}  \& {Martin}}{{Dove}
  et~al.}{1987}]{Dove:1987}
{Dove} J.~E.,  {Rusk} A.~C.~M.,  {Cribb} P.~H.,   {Martin} P.~G.,  1987, ApJ, 318, 379

\bibitem[\protect\citeauthoryear{{Draine} \& {Bertoldi}}{{Draine} \&
  {Bertoldi}}{1996}]{Drain:1996}
{Draine} B.~T.,  {Bertoldi} F.,  1996, ApJ, 468, 269

\bibitem[\protect\citeauthoryear{{Ferland}, {Peterson}, {Horne}, {Welsh}  \&
  {Nahar}}{{Ferland} et~al.}{1992}]{Ferland:1992}
{Ferland} G.~J.,  {Peterson} B.~M.,  {Horne} K.,  {Welsh} W.~F.,   {Nahar}
  S.~N.,  1992, ApJ, 387, 95

\bibitem[\protect\citeauthoryear{Flower et al.}{2000}]{Flower2000}
Flower, D. R., Le Bourlot, J., \& Pineau des For\^{e}ts, G. 2000, MNRAS, 314, 753

\bibitem[\protect\citeauthoryear{Fukushima et al.}{2018}]{Fukushima2018}
Fukushima, H., Omukai, K., \& Hosokawa, T. 2018, MNRAS, 473, 4754

\bibitem[\protect\citeauthoryear{Galli \& Palla}{1998}]{Galli1998}
Galli, D., \& Palla, F. 1998, A\&A, 335, 403

\bibitem[\protect\citeauthoryear{Gammie}{2001}]{Gammie2001}
Gammie, C. F. 2001, ApJ, 553, 174

\bibitem[\protect\citeauthoryear{Gerlich}{1982}]{Gerlich1982}
Gerlich, D. 1982, in Lindinger W., Howorka F., M\"{a}rk T. D., eds, 
  Symposium on Atomic and Surface Physics. Kluwer, Dordrecht, p. 304

\bibitem[\protect\citeauthoryear{{Glover}}{{Glover}}{2008}]{Glover:2008-3}
{Glover} S.,  2008, in {O'Shea} B.~W.,  {Heger} A.,  eds,  American Institute
  of Physics Conference Series Vol. 990, First Stars III. pp 25--29

\bibitem[\protect\citeauthoryear{Glover \& Abel}{2008}]{Glover:2008-8}
Glover, S. C. O., \& Abel, T. 2008, MNRAS, 388, 1627

\bibitem[\protect\citeauthoryear{Glover}{2015}]{Glover2015}
Glover, S. C. O. 2015, MNRAS, 453, 2901

\bibitem[\protect\citeauthoryear{Hollenbach \& Mckee}{1979}]{Hollenbach1979}
Hollenbach, D., \& Mckee, C. F. 1979, ApJS, 41, 555

\bibitem[\protect\citeauthoryear{Hollenbach \& Mckee}{1989}]{Hollenbach1989}
Hollenbach, D., \& Mckee, C. F. 1989, ApJ, 342, 306

\bibitem[\protect\citeauthoryear{Hosokawa et al.}{2013}]{Hosokawa2013}
Hosokawa T., Yorke H.W., Inayoshi K., Omukai K., Yoshida N., 2013, ApJ, 778, 178

\bibitem[\protect\citeauthoryear{{Janev}, {Langer}  \& {Evans}}{{Janev}
  et~al.}{1987}]{Janev:1987}
{Janev} R.~K.,  {Langer} W.~D.,   {Evans} K.,  1987, {Elementary processes in
  Hydrogen-Helium plasmas - Cross sections and reaction rate coefficients}.
Springer Verlag Berlin Heidelberg

\bibitem[\protect\citeauthoryear{Johnson \& Gammie}{2003}]{Gammie2003}
Johnson, B. M., \& Gammie, C. F. 2003, ApJ, 597, 131

\bibitem[\protect\citeauthoryear{Joos et al.}{2013}]{Joos2013}
Joos, M., Hennebelle, P., Ciardi, A., Fromang, S. 2013, A\&A, 554, 17

\bibitem[\protect\citeauthoryear{Kadam et al.}{2019}]{Kadam2019}
Kadam, K., Vorobyov, E. I., Regaly, Z., K\'osp\'al, A., \& \'Abrah\'am, P. 2019, ApJ, 882, 96

\bibitem[\protect\citeauthoryear{Kimura}{2016}]{Kimura2016}
Kimura, S. S., Kunitomo, M., \& Takahashi, S. Z. 2016, MNRAS, 461
2257

\bibitem[\protect\citeauthoryear{Kley}{1999}]{Kley1999}
Kley, W., 1999, MNRAS 303, 696

\bibitem[\protect\citeauthoryear{{Kreckel}, {Bruhns}, {{\v C}{\'{\i}}{\v z}ek},
  {Glover}, {Miller}, {Urbain}  \& {Savin}}{{Kreckel}
  et~al.}{2010}]{Kreckel:2010}
{Kreckel} H.,  {Bruhns} H.,  {{\v C}{\'{\i}}{\v z}ek} M.,  {Glover} S.~C.~O.,
  {Miller} K.~A.,  {Urbain} X.,   {Savin} D.~W.,  2010, Science, 329, 69

\bibitem[\protect\citeauthoryear{{Lenzuni}, {Chernoff}  \&
  {Salpeter}}{{Lenzuni} et~al.}{1991}]{Lenzuni:1991}
{Lenzuni} P.,  {Chernoff} D.~F.,   {Salpeter} E.~E.,  1991, ApJS, 76, 759

\bibitem[\protect\citeauthoryear{Machida et al.}{2010}]{Machida2010}
Machida, M. N., Inutsuka, S., and Matsumoto, T. 2010, 724, 1006

\bibitem[\protect\citeauthoryear{Matsukoba et al.}{2019}]{Matsukoba2019}
Matsukoba, R., Takahashi, S.~Z., Sugimura, K., and Omukai, K. 2019, MNRAS, 484, 2605

\bibitem[\protect\citeauthoryear{Mayer et al.}{2007}]{Mayer2007}
Mayer, L., Lufkin, G., Quinn, T., \& Wadsley, J. 2007, ApJL, 661,  77

\bibitem[\protect\citeauthoryear{Mayer \& Duschl}{2005}]{Mayer2005}
Mayer, M., \& Duschl, W. J. 2005, MNRAS, 358, 614

\bibitem[\protect\citeauthoryear{Mercer \& Stamatellos}{2017}]{Mercer2017}
Mercer, A., \& Stamatellos, D. 2017, MNRAS, 465, 2

\bibitem[\protect\citeauthoryear{Meru \& Bate}{2011}]{Meru2011}
Meru F., Bate M. R., 2011, MNRAS, 411, L1

\bibitem[\protect\citeauthoryear{Meru}{2015}]{Meru2015}
Meru, F. 2015 MNRAS, 454, 2529

\bibitem[\protect\citeauthoryear{M\"ueller et al.}{2012}]{Mueller2012}
M\"uller, T. W. A., Kley, W., Meru, F. 2012, A\&A, 541, 123

\bibitem[\protect\citeauthoryear{Nayakshin}{2017}]{Nayakshin2017}
Nayakshin, C. 2017, PASA, 34, 2

\bibitem[\protect\citeauthoryear{Omukai et al.}{2005}]{Omukai2005}
Omukai, K., Tsuribe, T., Schneider, R., \& Ferrara, A., 2005, ApJ, 626, 627

\bibitem[\protect\citeauthoryear{Omukai et al.}{2010}]{Omukai2010}
Omukai, K., Hosokawa, T., \& Yoshida, N., 2010, ApJ, 722, 1793

\bibitem[\protect\citeauthoryear{Oya et al.}{2016}]{Oya2016}
Oya, Y., Sakai, N., López-Sepulcre, A., et al. ApJ, 824, 88

\bibitem[\protect\citeauthoryear{{Palla}, {Salpeter}  \& {Stahler}}{{Palla}
  et~al.}{1983}]{Palla:1983}
{Palla} F.,  {Salpeter} E.~E.,   {Stahler} S.~W.,  1983, ApJ, 271, 632

\bibitem[\protect\citeauthoryear{Pavlyuchenkov et al.}{2015}]{Pavlyuchenkov2015}
Pavlyuchenkov Y. N., Zhilkin A. G., Vorobyov E. I., Fateeva
A. M., 2015, Astron. Rep., 59, 133

\bibitem[\protect\citeauthoryear{Pollack et al.}{1994}]{Pollack1994}
Pollack, J. B., Hollenbach, D., Beckwith, S., Simonelli, D. P., Roush, T., \& Fong, W. 1994, ApJ, 421, 615

\bibitem[\protect\citeauthoryear{Pringle}{1981}]{Pringle1981}
Pringle, J. E., 1981, ARA\&A, 19, 137.

\bibitem[\protect\citeauthoryear{Rice et al.}{2003}]{Rice2003}
Rice W. K. M., Armitage P. J., Bate M. R., \& Bonnell I. A., 2003, MNRAS, 339, 1025

\bibitem[\protect\citeauthoryear{Rice \& Armitage}{2009}]{Rice2009}
Rice, W. K. M. \& Armitage, P. J.  2009, MNRAS, 396, 2228

\bibitem[\protect\citeauthoryear{Rice et al.}{2010}]{Rice2010}
Rice, W. K. M.,   Mayo, J. H., \& Armitage, P. J. 2010, MNRAS, 402, 1740

\bibitem[\protect\citeauthoryear{Rice \& Nayakshin}{2018}]{Rice2018}
Rice, K., \& Nayakshin, S. 2018, MNRAS, 475, 921

\bibitem[\protect\citeauthoryear{Savin}{2002}]{Savin2002}
Savin, D. W. 2002, ApJ, 566, 599

\bibitem[\protect\citeauthoryear{Seifried et al.}{2013}]{Seifried2013}
Seifried, D., Banerjee, R., Pudritz, R. E., \&  Klessen, R. S. 2013, MNRAS, 432, 3320

\bibitem[\protect\citeauthoryear{Semenov et al.}{2003}]{Semenov2003}
Semenov, D., Henning, T., Helling, C., Ilgner, M., \& Sedlmayr, E. 2003, A\&A, 410, 611

\bibitem[\protect\citeauthoryear{Shavitt}{1959}]{Shavitt1959}
Shavitt, I. 1959, J. Chem. Phys., 31, 1359

\bibitem[\protect\citeauthoryear{Stamatellos et al.}{2007}]{Stamatellos2007}
Stamatellos, D., Whitworth, A. P., Bisbas, T., \&  Goodwin, S. 2007, A\&A, 475, 37

\bibitem[\protect\citeauthoryear{Stone \& Norman}{1992}]{SN1992}
Stone, J. M., \& Norman, M. L. 1992, ApJS, 80, 753

\bibitem[\protect\citeauthoryear{Tanaka \& Omukai}{2014}]{Tanaka2014}
Tanaka, K. E. I., \& Omukai, K. 2014, MNRAS, 439, 1884

\bibitem[\protect\citeauthoryear{Tielens \& Hollenbach}{1985}]{Tielens1985}
Tielens, A. G. G. M., \& Hollenbach, D. J. 1985, ApJ, 291, 722

\bibitem[\protect\citeauthoryear{{Trevisan} \& {Tennyson}}{{Trevisan} \&
  {Tennyson}}{2002}]{Trevisan:2002}
{Trevisan} C.~S.,  {Tennyson} J.,  2002, Plasma Physics and Controlled Fusion, 44, 1263

\bibitem[\protect\citeauthoryear{Truelove et al.}{1998}]{Truelove1998}
Truelove, J. K., Klein, R. I., McKee, C. F., et al. 1998, ApJ, 495, 821

\bibitem[\protect\citeauthoryear{Tsukamoto et al.}{2015}]{Tsukamoto2015}
Tsukamoto, Y., Takahashi, S. Z., Machida, M. N., \& Inut-
suka, S. 2015c, MNRAS, 446, 1175

\bibitem[\protect\citeauthoryear{van't Hoff et al.}{2020}]{Hoff2020}
van't Hoff, Merel L. R., van Dishoeck, E. F., Jørgensen, J. K., Calcutt, Hannah 2020, A\&A, 633, 7

\bibitem[\protect\citeauthoryear{Vorobyov}{2010}]{Vorobyov2010}
Vorobyov, E. I. 2010, ApJ, 713, 1059

\bibitem[\protect\citeauthoryear{Vorobyov \& Basu}{2009}]{VB2009}
Vorobyov, E. I., \& Basu, S. 2009, ApJ, MNRAS, 393, 822

\bibitem[\protect\citeauthoryear{Vorobyov \& Basu}{2010}]{VB2010}
Vorobyov, E. I., \& Basu, S. 2010, ApJ, 719, 1896

\bibitem[\protect\citeauthoryear{Vorobyov}{2013}]{Vorobyov2013}
Vorobyov, E. I. 2013, A\&A, 552, 129

\bibitem[\protect\citeauthoryear{Vorobyov et al.}{2018}]{Vorobyov2018}
Vorobyov, E. I., Akimkin, V., Stoyanovskaya, O.,
Pavlyuchenkov, Y., \& Liu, H. B. 2018, A\&A, 614, A98.

\bibitem[\protect\citeauthoryear{Vorobyov et al.}{2019}]{Vorobyov2019}
Vorobyov, E. I., Skliarevskii, A. M., Elbakyan, V. G., Pavlyuchenkov, Y., Akimkin, V., Guedel, M. 2019, A\&A, 657, 124.

\bibitem[\protect\citeauthoryear{{Wishart}}{{Wishart}}{1979}]{Wishart:1979}
{Wishart} A.~W.,  1979, MNRAS, 187, 59P

\bibitem[\protect\citeauthoryear{Wurster \& Bate}{2019}]{Wurster2019}
Wurster, J., \& Bate, M. R. 2019, MNRAS 486, 2587

\bibitem[\protect\citeauthoryear{Yorke \& Bodenheimer}{2008}]{Yorke2008}
Yorke H. W., \& Bodenheimer P., 2008, in Beuther H., Linz H., Henning T.,
eds, ASP Conf. Ser. Vol. 387, Massive Star Formation: Observations
Confront Theory. Astron. Soc. Pac., San Francisco, p. 189

\bibitem[\protect\citeauthoryear{Zhang \& Zhu}{2020}]{Zhang2020}
Zhang, S., \& Zhu, Z. 2020, MNRAS, 493, 2287

\bibitem[\protect\citeauthoryear{Zhu et al.}{2012}]{Zhu2012}
Zhu, Z., Hartmann, L., Nelson, R. P., \& Gammie, C. F. 2012, ApJ, 746, 110

\bibitem[\protect\citeauthoryear{Ziampras et al.}{2020}]{Ziampras2020}
Ziampras, A. Kley, W., \& Dullemond, C. P. 2020,  arXiv:2003.02298

\end{thebibliography}
\end{document}